\documentclass[12pt,preprint]{aastex}
\usepackage{natbib}

\shorttitle{2011 $\delta$ Scorpii Periastron Passage}
\shortauthors{A.S. Miroshnichenko et al.}

\begin{document}

\title{The 2011 Periastron Passage of the Be Binary $\delta$ Scorpii}

\author{A.~S.~Miroshnichenko$^1$}
\affil{$^1$Department of Physics and Astronomy, University of North
Carolina at Greensboro, Greensboro, NC 27402--6170, USA}

\author{A.~V.~Pasechnik$^2$}
\affil{$^2$Tuorla Observatory, Department of Physics and Astronomy,
University of Turku, Finland}

\author{N.~Manset$^3$}
\affil{$^3$CFHT Corporation, 65--1238 Mamalahoa Hwy, Kamuela, HI
96743}

\author{A.~C.~Carciofi$^4$}
\affil{$^4$Instituto de Astronomia, Geof\'isica e Ci\'encias
Atmosf\'ericas, Universidade de S\~{a}o Paulo, Brazil}

\author{Th.~Rivinius$^5$}
\affil{$^5$European Southern Observatory, Chile}

\author{S.~\v{S}tefl$^{5,6}$} \affil{$^5$European Southern
Observatory,Chile} \affil{$^6$Joint ALMA Observatory, Chile}

\author{V.~V.~Gvaramadze$^{7,8}$}
\affil{$^{7}$Sternberg Astronomical Institute, Lomonosov Moscow
State University, Universitetskij Pr. 13, Moscow 119992, Russia}
\affil{$^{8}$Isaac Newton Institute of Chile, Moscow Branch,
Universitetskij Pr. 13, Moscow 119992, Russia}

\author{J.~Ribeiro$^9$}
\affil{$^9$Observatorio do Instituto Geografico do Exercito, Lisboa,
Portugal}

\author{A.~Fernando$^{10}$}
\affil{$^{10}$ATALAIA.org Group---Lisboa, Portugal}

\author{T.~Garrel$^{11}$}
\affil{$^{11}$Observatoire de Foncaude, Montpellier, France}

\author{J.~H.~Knapen$^{12,13}$}
\affil{$^{12}$Instituto de Astrofísica de Canarias, E--38205 La
Laguna, Tenerife, Spain} \affil{$^{13}$Departamento de Astrof\'\i
sica, Universidad de La Laguna, 38206 La Laguna, Spain}

\author{C.~Buil$^{14}$}
\affil{$^{14}$Castanet Tolosan Observatory, Toulouse, France}

\author{B.~Heathcote$^{15}$}
\affil{$^{15}$Barfold Observatory, Glenhope, Victoria 3444,
Australia}

\author{E.~Pollmann$^{16}$}
\affil{$^{16}$Emil-Nolde-Str. 12, 51375, Leverkusen, Germany}

\author{B.~Mauclaire$^{17}$}
\affil{$^{17}$Observatoire du Val d$^{\prime}$Arc, France}

\author{O.~Thizy$^{18}$}
\affil{$^{18}$Shelyak Instruments, France}

\author{J~.Martin$^{19}$}
\affil{$^{19}$Barber Research Observatory, Department of Physics and
  Astronomy, University of Illinois-Springfield, IL 62703, USA}

\author{S.~V.~Zharikov$^{20}$}
\affil{$^{20}$Instituto de Astronomia, Universidad Nacional
Aut\'onoma de Mexico, Apdo. Postal 877, Ensenada, 22800, Baja
California, Mexico}

\author{A.~T.~Okazaki$^{21}$}
\affil{$^{21}$Faculty of Engineering, Hokkai-Gakuen University,
Toyohira-ku, Sapporo 062--8605, Japan}

\author{T.~L.~Gandet$^{22}$}
\affil{$^{22}$Lizard Hollow Observatory, P.O. Box 89175, Tucson, AZ
  85752-9175, USA}

\author{T.~Eversberg$^{23}$}
\affil{$^{23}$Schn\"orringen Telescope Science Institute,
Waldbr\"ol, Germany}

\author{N.~Reinecke$^{24}$}
\affil{$^{24}$Bonn, Germany}

\altaffiltext{1}{This paper is partially based on observations
obtained at the Canada-France-Hawaii Telescope (CFHT) which is
operated by the National Research Council of Canada, the Institut
National des Sciences de l$^{\prime}$Univers of the Centre National
de la Recherche Scientifique de France, and the University of
Hawaii, the 2.2m MPG telescope operated at ESO/La Silla under
program IDs 086.A--9019 and 087.A--9005, the IAC80 telescope in the
Spanish Observatorio del Teide of the Instituto de Astrofi\'\i sica
de Canarias, and data from the ELODIE archive at the Observatoire de
Haute-Provence. }

\begin{abstract}
We describe the results of the world-wide observing campaign of the
highly eccentric Be binary system $\delta$ Scorpii 2011 periastron
passage which involved professional and amateur astronomers. Our
spectroscopic observations provided a precise measurement of the
system orbital period at $10.8092\pm 0.0005$ years. Fitting of the
He {\sc ii} 4686 \AA\ line radial velocity curve determined the
periastron passage time on 2011 July 3, UT 9:20 with a 0.9--day
uncertainty. Both these results are in a very good agreement with
recent findings from interferometry. We also derived new
evolutionary masses of the binary components (13 and 8.2 M$_\odot$)
and a new distance of 136 pc from the Sun, consistent with the
HIPPARCOS parallax. The radial velocity and profile variations
observed in the H$\alpha$ line near the 2011 periastron reflected
the interaction of the secondary component and the circumstellar
disk around the primary component. Using these data, we estimated a
disk radius of 150 R$_\odot$. Our analysis of the radial velocity
variations measured during the periastron passage time in 2000 and
2011 along with those measured during the 20th century, the high
eccentricity of the system, and the presence of a bow shock-like
structure around it suggest that $\delta$ Sco might be a runaway
triple system. The third component should be external to the known
binary and move on an elliptical orbit that is tilted by at least
40$\degr$ with respect to the binary orbital plane for such a system
to be stable and responsible for the observed long-term radial
velocity variations.
\end{abstract}


\section{Introduction} \label{intro}

$\delta$ Scorpii is one of the brightest stars in the sky. Until
2000, its visual brightness was $V$ = 2.32 mag; since then, and due
to its transition to a Be phase \citep{Carciofi2006}, it has been
even brighter ($V \le$ 1.6 mag). It was resolved interferometrically
into two components in the 1970's, and the observations indicated a
very eccentric orbit with a period of $\sim$ 10.6 years
\citep{Hartkopf1996}. The system is not eclipsing, and the secondary
component is $\Delta B = 1.78\pm0.03$ mag fainter than the primary
one \citep{Tango2009}.

$\delta$ Sco was used as a spectral classification standard (B0 {\sc
iv}) during the entire 20th century. The first signs of weak
emission in the H$\alpha$ line were detected by \citet{Cote1993}
close to a periastron time in 1990, but the emission only slightly
varied throughout the next orbital cycle
\citep{Miroshnichenko2001,Koubsky2005}. In June 2000, near the
following periastron time, the system was found to be $\sim$0.03 mag
brighter than usual, while a spectrum taken shortly after revealed a
noticeable H$\alpha$ emission line \citep{Fabregat2000}. A follow up
spectroscopic campaign resulted in the detection of a growing line
emission and a significant variation of the radial velocity
\citep{Miroshnichenko2001}. These data constrained the periastron
time (2000 September $9\pm3$) which was predicted to occur a few
months earlier from the interferometric data. Since that time the
circumstellar (CS) disk around the primary component of $\delta$~Sco
grew larger, the system brightness varied between $V \sim$ 1.6 and
2.3 mag, but the line emission has never disappeared. The disk
development has been documented in \citet{Miroshnichenko2003} and
\citet{Carciofi2006}.

The erroneous interferometric prediction of the periastron time in
2000 reflected the accuracy of this technique at that time. The
orbital separation of the components that varies from 6 to 200\,mas
only allowed to resolve them during about half the orbital cycle,
mostly around apastron. Recent advances in this field made it
possible to cover the entire orbit
\citep[e.g.,][]{Tycner2011,Meilland2011}. The interferometric
observations obtained since 2000 combined with the orbit derived
from the radial velocity curve \citep{Miroshnichenko2001} resulted
in predicting the periastron time in 2011 with a high accuracy
\citep[2011 July 2--6,][]{Tango2009,Tycner2011,Meilland2011}.
Moreover, interferometry obtained in July 2011
\citep{Che2012,Stefl2012}, right after the periastron, confirmed the
prediction.

Spectroscopy is another powerful tool in studying binaries that is
capable of verifying interferometric results and obtaining
additional information about objects' properties. In order to get
ready for the 2011 periastron, a spectroscopic observing campaign
was planned well in advance \citep[e.g.,][]{Miroshnichenko2009}. Its
important feature was a broad participation of amateur astronomers
whose involvement in spectroscopic observations of various objects
significantly increased over the last decade
\citep[e.g.,][]{Fahed2011}.

This paper describes the main results of the $\delta$ Sco 2011
periastron spectroscopic observing campaign as well as some
additional findings that suggest new ideas about the system's nature
and evolution. The observations are described in Sect.
\ref{observations}, analysis of the 2011 periastron radial velocity
curves and the orbital period determination are presented in Sect.
\ref{analysis}, historical radial velocity data along with new ideas
about the system's nature are discussed in Sect. \ref{discussion},
and conclusions and predictions are summarized in Sect.
\ref{conclusions}.

\section{Observations}\label{observations}

Spectroscopic monitoring of $\delta$ Sco began shortly after its
brightening discovery in 2000 \citep{Fabregat2000} and has been
ongoing since that time. Due to its brightness, it has mostly been
observed with under 1m size telescopes. The amateur community
participation has been steadily growing towards the time of the 2011
periastron. While only a few spectra of $\delta$ Sco were obtained
by amateurs in 2000, this number grew to 200 spectra in 83 nights in
2010 and over 300 spectra in 149 nights in 2011. The amateurs
contribution in 2010--2011 is comparable with that of professional
astronomers in the number of spectra and exceeds it in the time
coverage.

The main goals of the 2011 campaign were 1) to obtain an accurate
radial velocity curve to independently constrain the orbital period,
and 2) to study line profile variations to search for effects of the
interaction between the disk and the secondary component as well as
possibly get information about the latter. The interaction was
expected due to a very small distance between the components at
periastron \citep[$\sim$15 radii of the primary
component,][]{Miroshnichenko2001} and the disk size comparable with
this distance \citep{Carciofi2006,Millan2010}.

Since not all of the involved observers were able to cover a wide
spectral range, it was suggested to focus on the two following
regions. First, a region around the H$\alpha$ emission line whose
profile is indicative of the tidal interaction due to a possible
disk restructuring. Second, a region around the photospheric He {\sc
ii} 4686 \AA\ line whose profile seems to be the least affected by
the disk material and is therefore the best tracer of the orbital
motion of the primary component. The He {\sc i} lines that are
abundant in the object's spectrum were rejected because their
profiles are subject to complicated distortions by the disk
material, and the CS contribution to them is difficult to quantify.

The main instruments that contributed to the campaign from the
professional side were the 3.6\,m CFHT with the spectropolarimeter
ESPaDOnS \citep{Donati1997}, which covers a region
$\sim$3600--10\,500\,\AA\ with a spectral resolving power $R
\sim70\,000$, and the 2.2\,m MPG telescope at ESO/La Silla with the
FEROS spectrograph \citep{Kaufer2000}, which covers a region
$\sim$3600--9200\,\AA\ with $R = 48\,000$. Over 300 individual
spectra of $\delta$ Sco were obtained at these facilities during 40
nights in 2011.

An important feature of the campaign was a ten night observing run
at the 0.8\,m IAC80 telescope of the Instituto de Astrof\'{i}sica de
Canarias at the Teide Observatory on the Canary Island of Tenerife
in Spain. The run took place between 2011 June 28 and 2011 July 8,
was centered on the periastron time predicted by \citet{Tycner2011},
and accomplished by a team of five authors of this paper (A.
Miroshnichenko, J. Ribeiro, A. Fernando, T. Garrel, and J. Knapen).
We used a private Lhires {\sc iii}
spectrograph\footnote{http://www.astrosurf.com/thizy/lhires3/index-en.html}
with $R \sim 17\,000$ in the H$\alpha$ region and $R \sim 21\,000$
in the He {\sc ii} 4686\,\AA\ line region. All nights of the run
were clear. We obtained over 100 individual exposures in the two
mentioned regions as well as in a region near H$\gamma$ and He {\sc
i} 4471\,\AA\ line.

The amateur community contribution to the campaign involved nearly
20 observers from France, Germany, Australia, Portugal, Spain, and
the USA. They used 0.2\,m to 0.4\,m telescopes with either \'echelle
or long-slit spectrographs with a range of $R = 1000$--22\,000. Only
the spectra with $R \ge10\,000$ were used in our analysis.

Data reduction was performed using Libre-ESpRIT data reduction
package \citep{Donati1997} for the ESPaDOnS data and ESO--MIDAS
package for the FEROS data. IRAF was used to reduce some of the
Tenerife campaign data, while most amateurs data were reduced with
software packages developed for amateur spectrographs, such as
Audela\footnote{http://www.audela.org/dokuwiki/doku.php/en/start}
and
IRIS\footnote{http://www.astrosurf.com/buil/isis/isis$\_$en.htm}.
Wavelength calibration was controlled by measuring positions of
telluric and interstellar (e.g., Na\,{\sc i} D$_{1,2}$) lines and
contemporaneous observations of radial velocity standards (e.g.,
$\alpha$\,Ser and $\delta$\,Oph). Typical uncertainties of the
radial velocity measurements are below 1\,km\,s$^{-1}$ for the
ESPaDOnS and FEROS data and 2--4\,km\,s$^{-1}$ for the amateur data.

\section{Data Analysis}\label{analysis}

\subsection{The H$\alpha$ line} \label{rv_curve}

\begin{figure}[t]
\begin{tabular}{cc}
\resizebox{7.5cm}{!}{\includegraphics{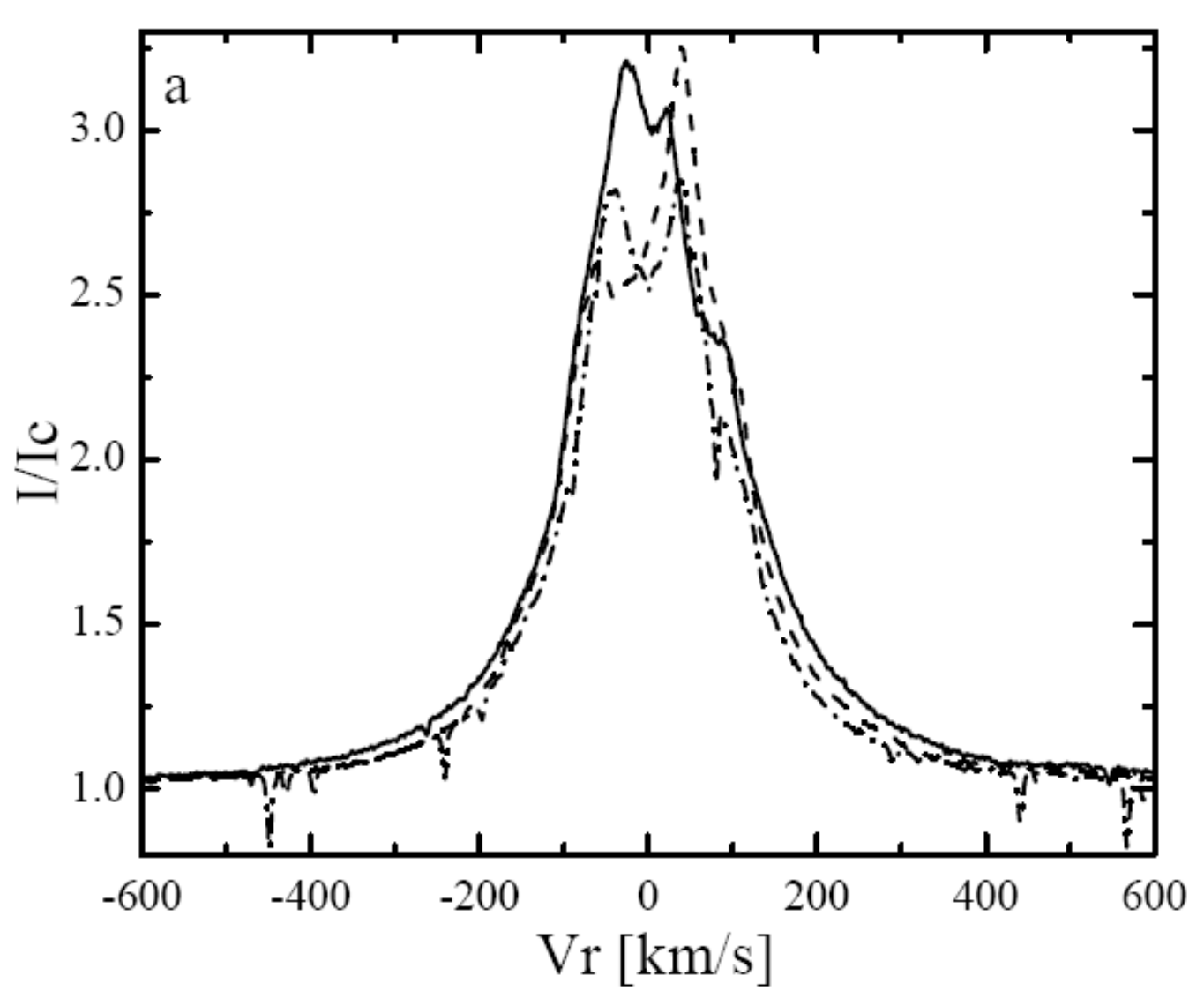}}&\resizebox{7.7cm}{!}{\includegraphics{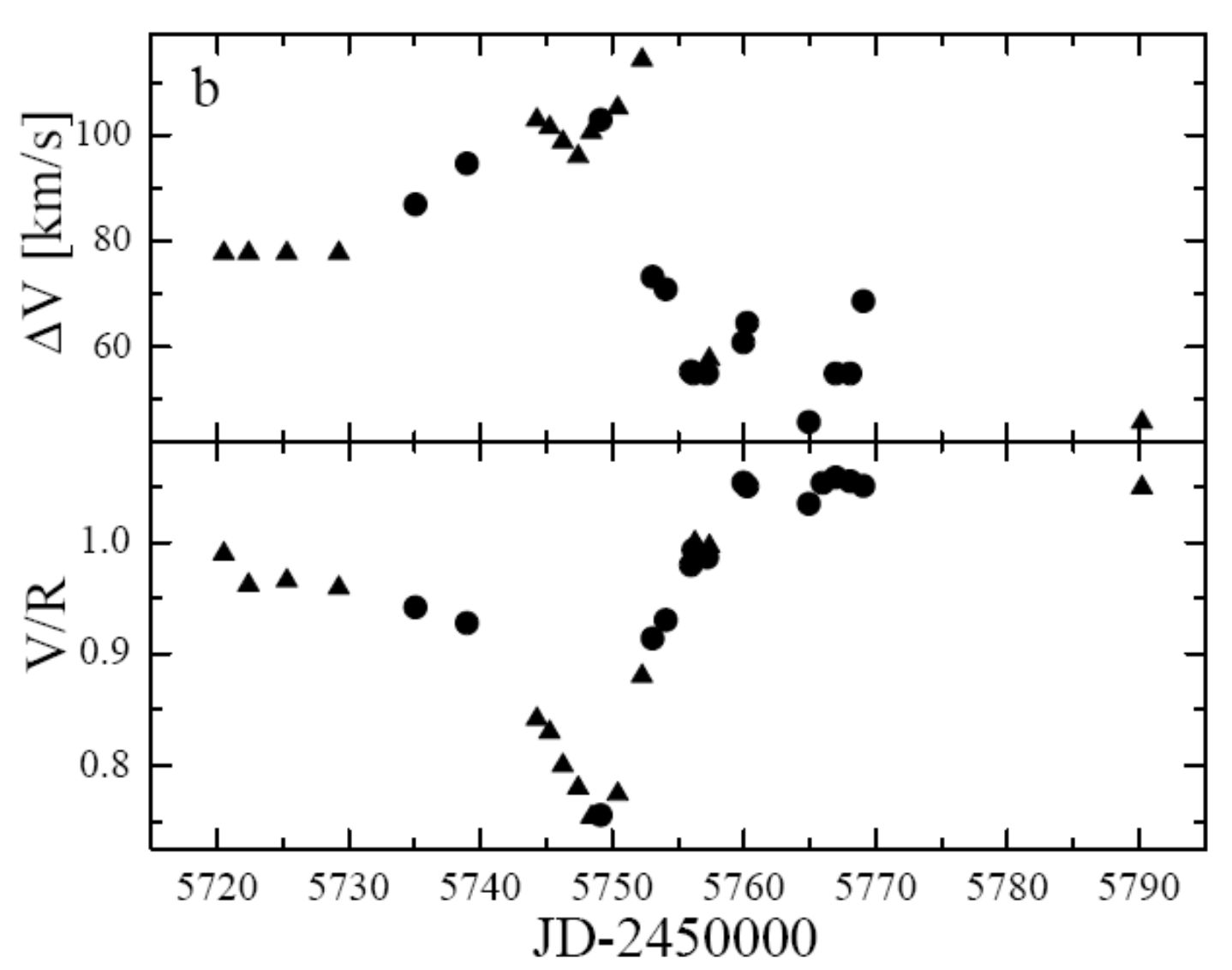}}\\
\resizebox{7.5cm}{!}{\includegraphics{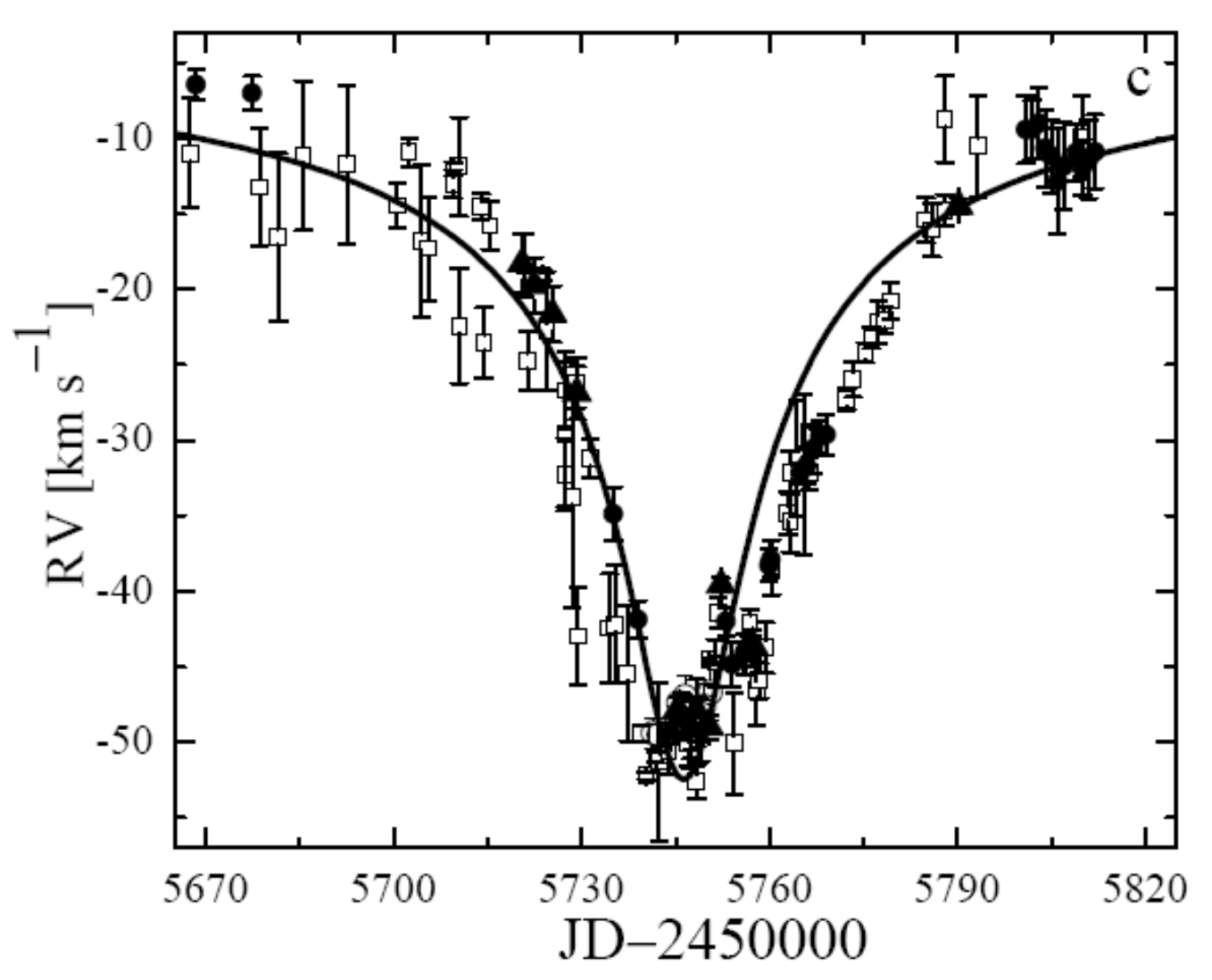}}&\resizebox{7.7cm}{!}{\includegraphics{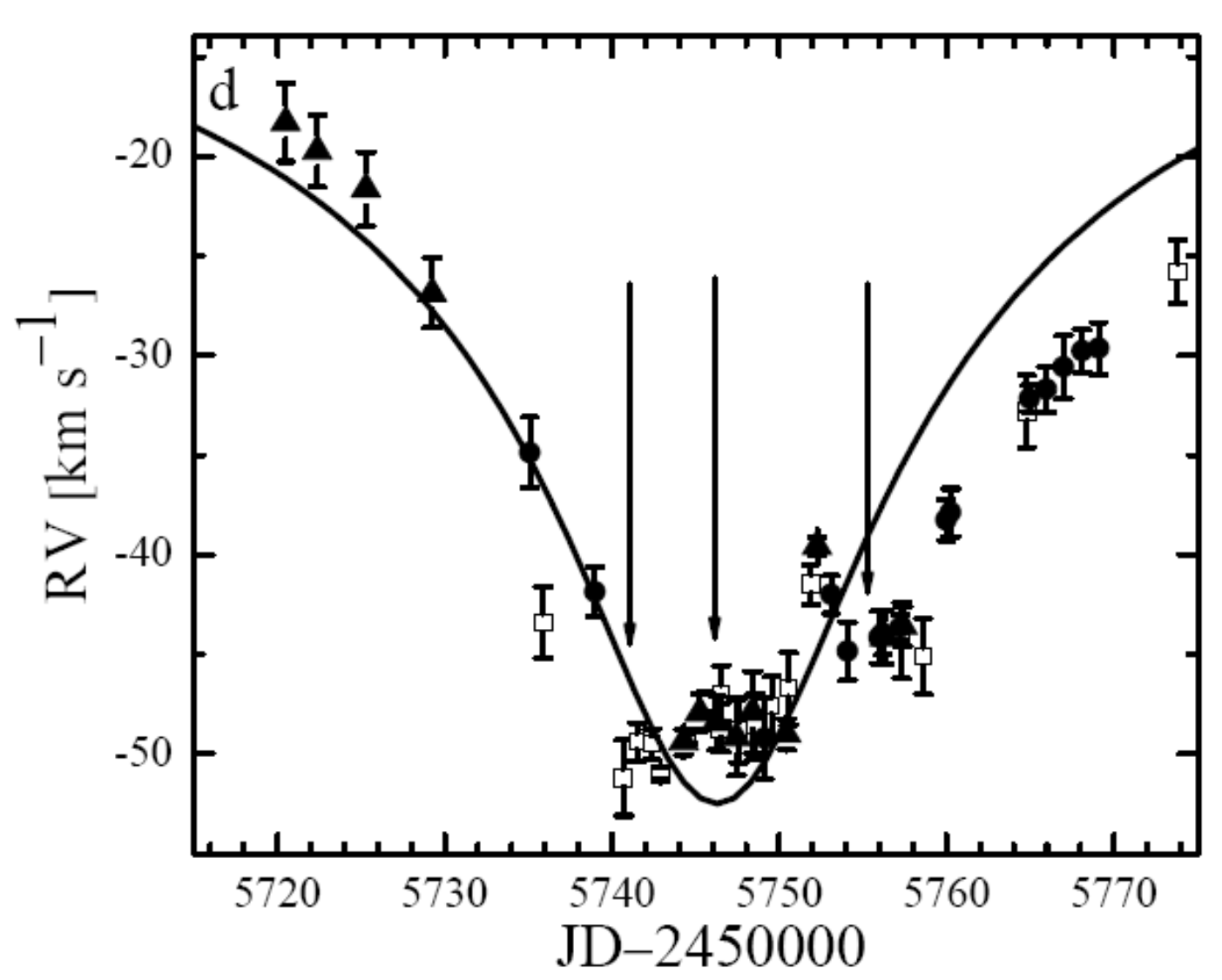}}\\
\end{tabular}
\caption{The profile and radial velocity variations of the H$\alpha$
line in the spectrum of $\delta$ Sco around the periastron time.
{\bf Panel a:} the line profile variations. CFHT spectra on three
dates are shown: 2011 June 8 (solid line), 2011 July 4 (dashed
line), and 2011 August 16 (dash-dotted line). The intensity is
normalized to the nearby continuum, the radial velocity is
heliocentric and shifted to the periastron time. No correction for
telluric lines was done. {\bf Panel b:} the line peak ratio and the
peak separation. Symbols: filled circles---CFHT/ESPaDOnS data and
filled triangles---ESO/FEROS data. {\bf Panel c:} all radial
velocity data obtained in March--October 2011. The heliocentric
radial velocities in km\,s$^{-1}$ are plotted against time in Julian
dates. The solid line represents the best fit radial velocity curve
to the data obtained during the 2000 periastron shifted by +6
km\,s$^{-1}$ \citep{Miroshnichenko2001}. Symbols: professional
data---the same as in panel b), open squares---all amateurs data.
{\bf Panel d:} the heliocentric radial velocities for a shorter
period of time near the periastron. Symbols are the same as in panel
c), but the amateurs data are averaged within 2--3 day periods. The
arrows mark three minima (see text). The central minimum roughly
coincides with the periastron time. \label{f1}}
\end{figure}

The double-peaked structure of the H$\alpha$ line profile was
resolved almost all the time in 2011 even at $R \sim$10\,000. The
peak intensity ratio was $V/R \ge 1$ until about a month before the
periastron time, then it reached a minimum at periastron, and came
back to the pre-periastron value about two weeks later (see Fig.
\ref{f1}ab). The line equivalent width increased from $\sim$11 \AA\
in May 2011 to $\sim$15 \AA\ in September, while the visual
continuum got brighter by $\sim$0.1 mag between late-June and a few
days after the periastron \citep{Rivinius2012}. Therefore the line
flux was rising as well. The latter is unlikely to be due to a
tidally induced enhanced mass loss from the primary, as the close
passage of the secondary only reduced the primary surface gravity by
$\sim 0.01$\%. Both the $V/R$ and the equivalent width changes began
virtually at the same time and might be explained by emission from
the disk material that was getting into the secondary Roche lobe.
The secondary was moving away from us until the periastron time and
might have caused the red peak enhancement.

The system brightness decreased after the periastron, and the above
mentioned further increase of the equivalent width was mostly due to
this effect. This is consistent with a constant disk contribution to
the $H$--band flux from 6 to 18 days after periastron reported by
\citet{Che2012}. We also note that in 2010 the H$\alpha$ radial
velocity was the same as before the 2000 periastron, but became
$\sim$ 10 km\,s$^{-1}$ more negative in January 2011, nearly two
months earlier than the radial velocity should have started to
change due to the binary components closeness. This effect might
have been due to some processes in the disk which are beyond the
scope of this paper. A more detailed analysis of the line profile
variations will be presented elsewhere (Rivinius et al., in prep.).

The radial velocity curve from the H$\alpha$ data is shown in Fig.
\ref{f1}cd. The measurements were accomplished using the mirrored
profile method \citep[see, e.g.,][]{Nemravova2012}. The measurement
results are presented in Table \ref{t1}. The curve shows a more
complicated structure compared to the one observed in 2000 and
expected from a binary system with no changes in the CS matter.
There were three turnovers marked by arrows in Fig. \ref{f1}d. The
first one occurred about five days before the periastron and was a
consequence of the line profile redshift ($V/R$ decrease) described
above. The second one occurred two days after the periastron, when
the $V/R$ ratio exhibited a minimum. The third one took place nearly
two weeks after the periastron on 2011 July 18 (JD2455760) and
coincided with the rapid change of the $V/R$ ratio back to
$\sim$1.1. After that time, the secondary with a part of the
circumprimary disk seemed to have moved beyond the disk boundary and
stopped affecting the remaining part of the disk.

The line profile variations near the periastron allow us to roughly
estimate the disk radius. The changes began to occur about 10 days
prior to the periastron, when the stars were $\sim$250 R$_\odot$
apart. Assuming an equilibrium situation, at this time the first
Lagrangian point between the two stars was laying $\sim$150
R$_\odot$ from the primary \citep{Eggleton1983}. This distance can
be regarded as an upper limit on the disk radius. It corresponds to
20 radii of the primary component assuming a distance of 136 pc (see
Sect. \ref{masses}). This is somewhat larger than the 104 R$_\odot$
for the semi-major FWHM of the primary H$\alpha$ disk measured by
\citet{Millan2010} in July 2007 during a brightness minimum period
when the system was as bright as before the disk began developing.

The system became $\sim$60\% optically brighter between 2009 and
2011. This brightening was very similar to the initial one in
2000/2001, but this time the disk was already present. The most
likely reason for the fading was clearing of the disk inner regions
due to the mass loss weakening \citep{Haubois2012}. However, the
disk outer regions must have survived, as the line emission was
observed all the time. Therefore, the circumprimary disk was
probably steadily growing during the entire orbital cycle. It is
beyond the scope of this paper to discuss this process in detail. We
only note that the disk size was $\sim$49 R$_\odot$ already in 2001
\citep{Carciofi2006}, not in 2005 as quoted by \citet{Meilland2011}.

The radial velocity of the H$\alpha$ line continued to change very
close to the binary model prediction since the end of July 2011
(solid line in Fig.\,\ref{f1}d). Overall, the effects of the
secondary close passage and its interaction with the disk made the
H$\alpha$ radial velocity curve broader than in 2000. Also, the
H$\alpha$ heliocentric radial velocity curve was shifted by $\sim$ 6
km\,s$^{-1}$ to the negative velocities compared to that in 2000
(Fig.\,\ref{f1}c). The shift may be due to a different disk density
distribution as well as to other possible reasons discussed in
Sect.\,\ref{discussion}.

\subsection{The He {\sc ii} 4686 \AA\ line}

\begin{figure}[!t]
\begin{tabular}{lll}
\resizebox{5.1cm}{!}{\includegraphics{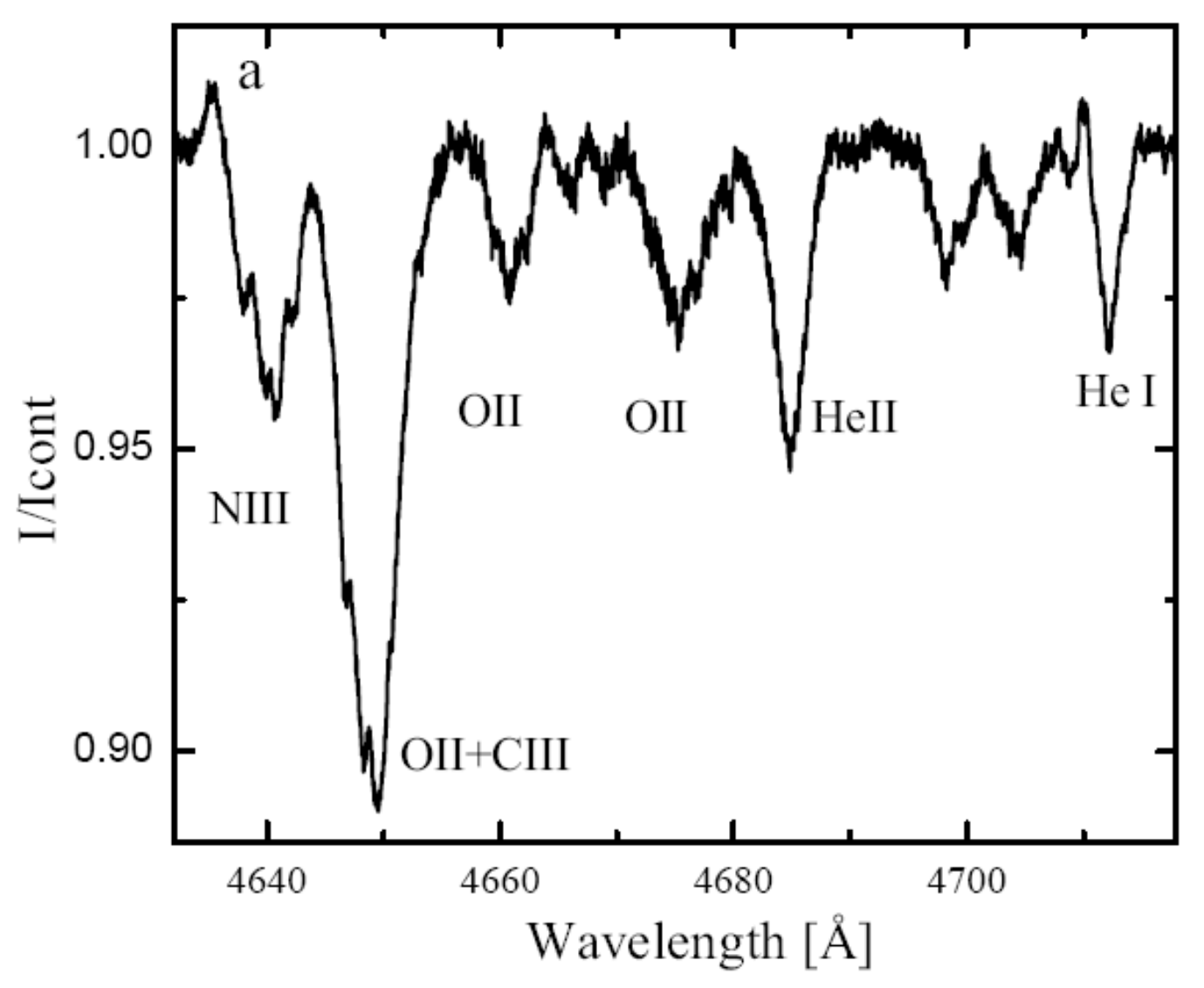}}&
\resizebox{5.1cm}{!}{\includegraphics{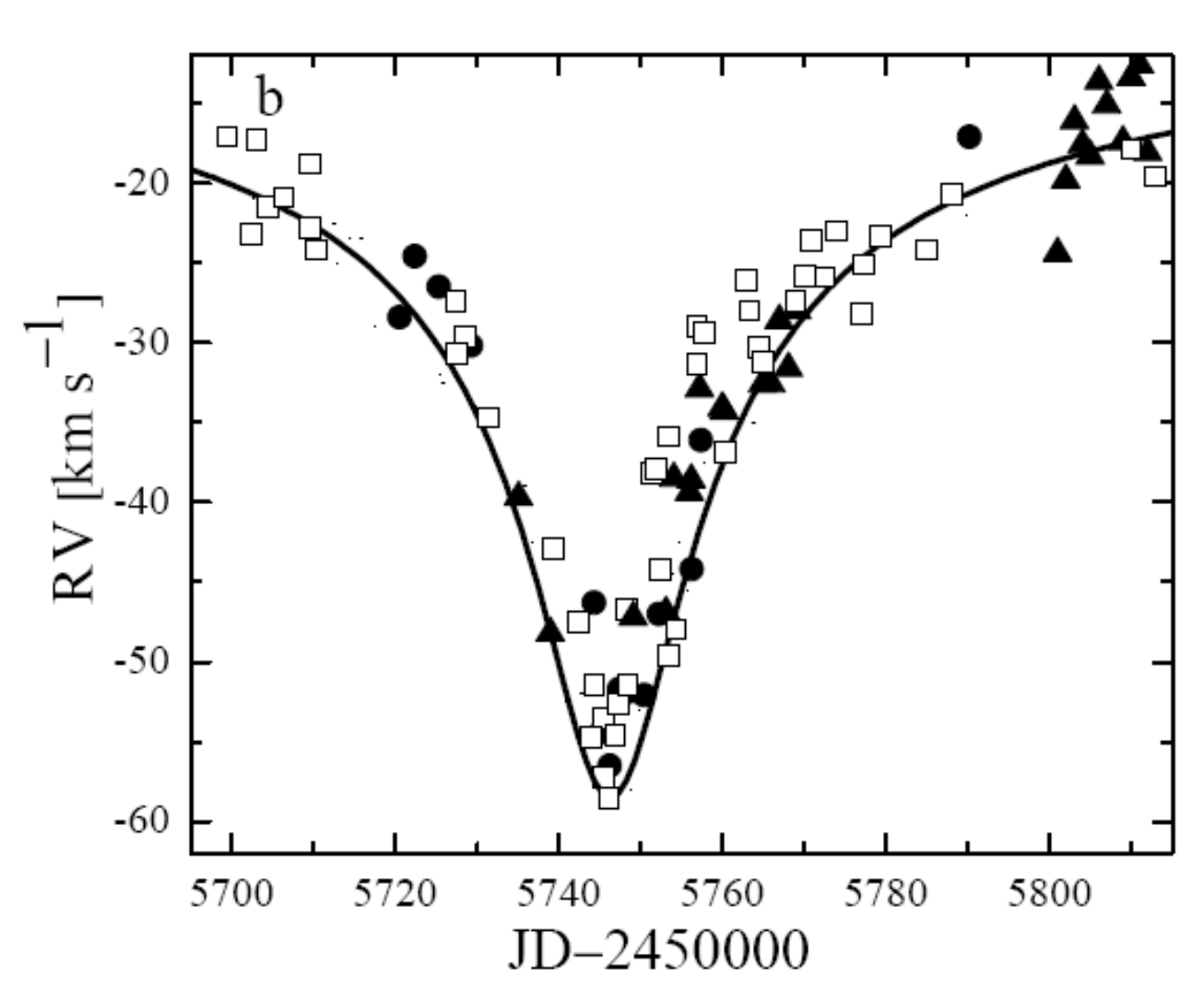}}&
\resizebox{5.1cm}{!}{\includegraphics{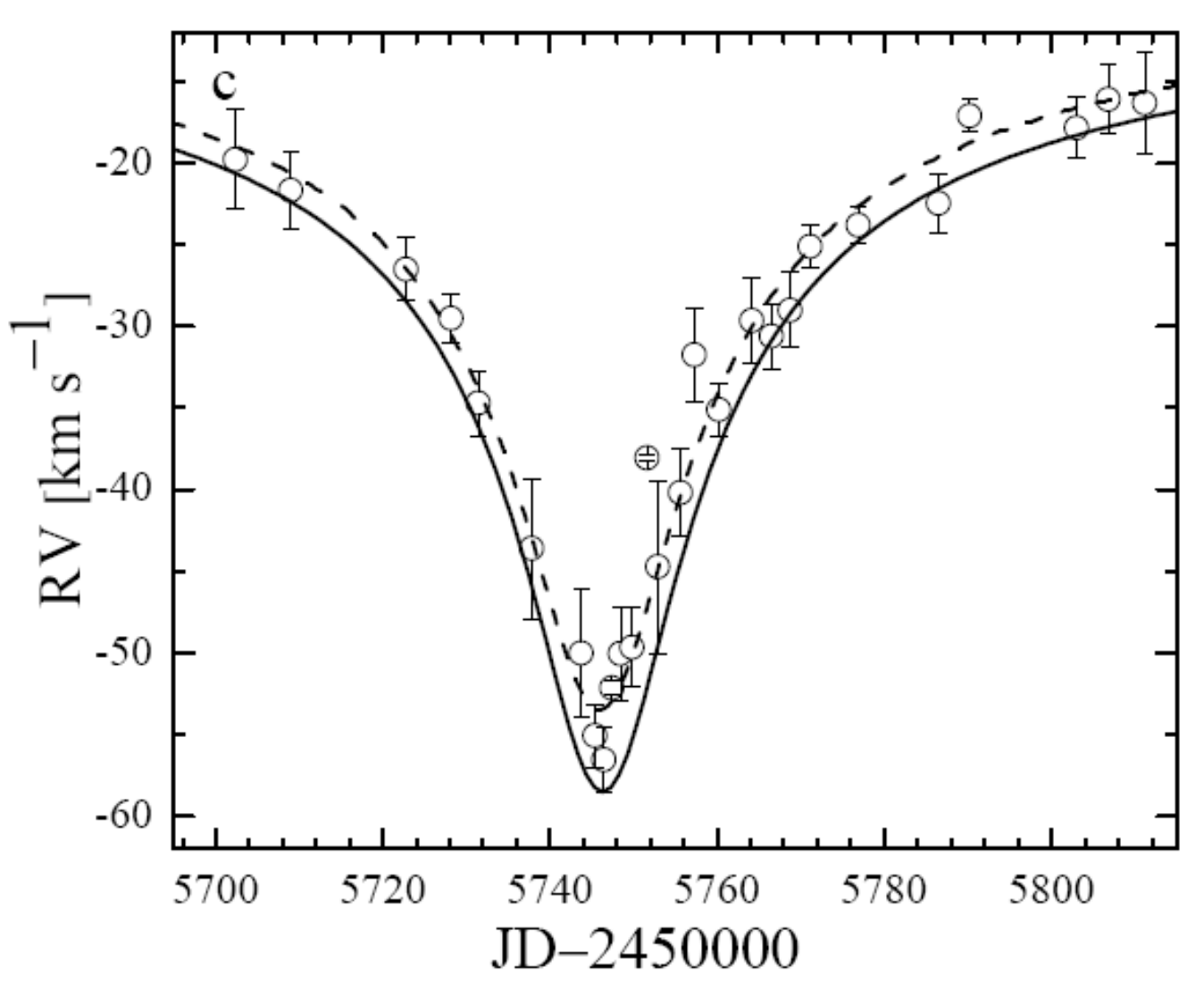}}\\
\end{tabular}
\caption{{\bf Panel a:} Part of the 2011 July 4 CFHT spectrum near
the He\,{\sc ii} 4686 \AA\ line. Intensities are normalized to the
underlying continuum, and heliocentric wavelengths are given in \AA.
{\bf Panel b:} Radial velocity data for the He {\sc ii} 4686 \AA\
line in the spectrum of $\delta$~Sco. The radial velocities and time
are given in the same units as in Fig.\,\ref{f1}. The solid line
represents the best fit heliocentric radial velocity curve to the
data obtained during the 2000 periastron from
\citet{Miroshnichenko2001}, no shift is applied. {\bf Panel c:} The
same set of radial velocities for the He {\sc ii} 4686 \AA\ line,
but averaged over 2--3 day periods to decrease the individual data
scatter. The professional and amateur data are mixed together on
this plot. The solid line is the same as in the Panel b). The dashed
line is the best fit for the averaged data from Table \ref{t3}.
\label{f2}}
\end{figure}

A spectral region that contains the He {\sc ii} 4686 \AA\ line is
shown in Fig.\,\ref{f2}a. The radial velocity of the He {\sc ii}
line was measured by fitting it to a Gaussian that for symmetric
lines gives very similar results to the mirrored profile method. The
measurement results are presented in Table \ref{t2}. We have 115
measurements of the He {\sc ii} line radial velocity between 2010
May 28 and 2012 October 12. Eighty seven of them (from 2011 May 18
and September 8) were used in the radial velocity curve fitting.
These include forty measurements from the ESPaDOnS and FEROS
spectra, four from IAC80, and forty three from the amateurs spectra.
The He {\sc ii} line profile kept stable, but its weakness
($\sim$0.95 of the continuum at the minimum intensity) seemed to be
the source of scatter seen in Fig.\,\ref{f2}b. In order to minimize
the scatter, we averaged the measurements within 2--3 day intervals
and fitted the resulting 25 data points to a binary orbital motion
model predictions.

The initial orbital parameters for the fitting were taken from
recent interferometry papers \citep[e.g.,][]{Meilland2011,Che2012}.
Most of them are very close to the ones obtained for the 2000 data
by \citet{Miroshnichenko2001}. Nevertheless, we carried out an
exhaustive least square fitting allowing all the orbital parameters
to vary. We also fitted both the 2000 H$\alpha$ and 2011 He {\sc ii}
data sets together to check for consistency of the orbital period
determination. Using data for these two lines in the same set is
legitimate because the disk in 2000 was very small and hardly
affected by the secondary component approach. Therefore, the
H$\alpha$ line very closely traced the motion of the primary
component at that time. The best fit parameters are shown in Table
\ref{t3}. They are a little different for the 2000 data set, because
this time we used the radial velocity semi-amplitude as a fit
parameter instead of the semi-major axis, the component's mass
ratio, and the orbital inclination angle. However, the updated
orbital elements are within the uncertainties given in
\citet{Miroshnichenko2001} and also consistent with the recent
interferometric solutions. The 2000 periastron epoch was refined
from September 9$\pm$3 to September 10.5$\pm$1.0.

The best fits to both the 2011 radial velocity data and the combined
2000 and 2011 data set show that the periastron occurred on 2011
July 3, 9:20 UT, with a 0.9--day uncertainty. Thus, the
interferometric result for the periastron epoch \citep{Che2012} was
confirmed spectroscopically with a shift of only $\sim$ 2 hours,
well within the uncertainties of both measurements. Our fitting also
constrained the orbital period at 3948.0$\pm$1.8 days, very close to
the latest interferometric solutions.

The parameter errors presented in Table \ref{t3} are larger than
those of the interferometric solution by \citet{Che2012}. The main
reason for this result is that the time period during which the
radial velocity changes significantly is much shorter (2--3 months)
than the system half orbital period during which interferometry data
were taken. Nevertheless, spectroscopy provides an independent
constraint for the radial velocity semi-amplitude which allows to
derive the components masses (see Sect. \ref{masses}).

The radial velocity curve derived for the He {\sc ii} line slightly
deviates from the one derived for the H$\alpha$ line in 2000
(Fig.\,\ref{f2}bc). It is slightly narrower and is shifted to the
positive velocities by $\sim 1.9$\,km\,s$^{-1}$. Although the
deviation seems small, it may indicate that the orbit is unstable.
We study this and other related observational features in Sect.
\ref{discussion}.

\subsection{Other Spectral Variations} \label{variations}

\begin{figure}
\begin{tabular}{cc}
\includegraphics[width=6.3cm,height=5cm]{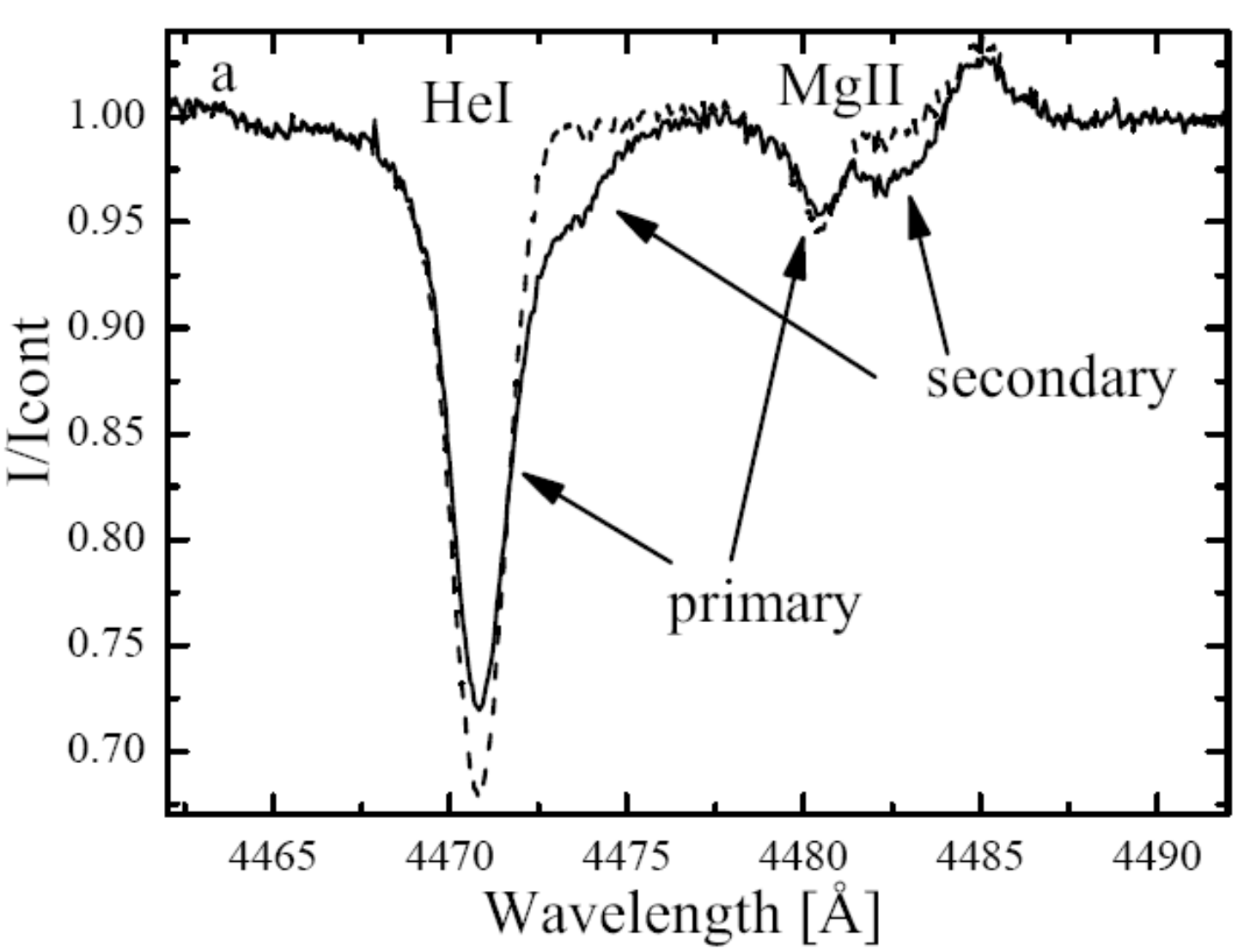}&
\includegraphics[width=6.3cm,height=5cm]{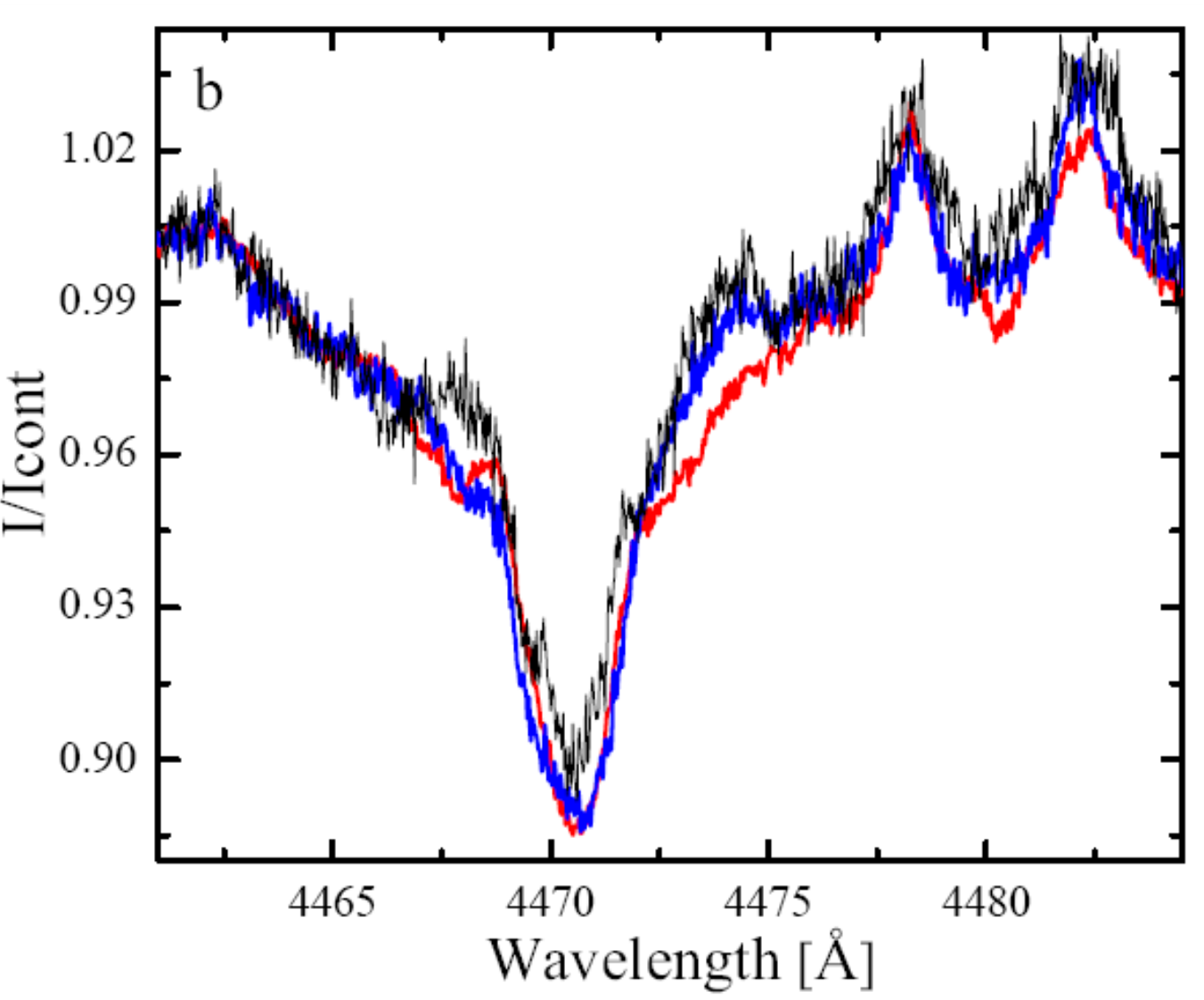}\\
\end{tabular}
\caption{Possible effects of the secondary component on the spectrum
of $\delta$ Sco at periastron. Panel a) Solid line shows a sum of
two spectra of BS\,1880, a B0.5\,V star (shown by a dashed line),
and BS\,801, a B3\,V star, scaled with a brightness difference of
$\Delta B =$ 1.78 mag. The components' radial velocity difference is
120 km\,s$^{-1}$. Panel b) shows the CFHT spectra taken before (blus
line, 2011 June 8), at (red line, 2011 July 4), and after the
periastron (black line, 2011 August 16). The Mg {\sc ii} 4482 \AA\
line has a double-peak emission profile. The wavelengths are shifted
to the periastron epoch. Intensities and wavelengths are in the same
units as in Fig.\,\ref{f2}b. \label{f3}}
\end{figure}

There was a hope to detect signs of the secondary component in the
spectrum due to a relatively large radial velocity difference at
periastron ($\sim$120 km\,s$^{-1}$). Some ideas about the secondary
parameters were deduced from the system's spectral energy
distribution, orbital elements, and the components brightness ratio
\citep[an early B--type dwarf, see discussion
in][]{Miroshnichenko2001,Miroshnichenko2009}. One of the expected
signatures of the secondary is shown in the left panel of Fig.
\ref{f3}. This prediction is based on a sum of two spectra taken
from the online archive of the Observatoire de Haute Provence (OHP)
obtained with the spectrograph
ELODIE\footnote{http://atlas.obs-hp.fr/elodie/} \citep[$R \sim$
48,000,][]{Moultaka2004}. We chose the B0.5 {\sc v} star BS\,1880 to
represent the primary and the B3 {\sc v} star BS\,801 to represent
the secondary. The sum was calculated assuming a brightness ratio of
$\Delta B$ = 1.78 mag (see Sect.\,\ref{intro}). No CS continuum
contribution was added to the composite spectrum, because we only
aimed at a qualitative demonstration of possible effects at
periastron.

Their projected rotational velocities listed in a recent catalog by
\citet{Glebocki2005} are 25--120 km\,s$^{-1}$ for BS\,1880 and
90--130 km\,s$^{-1}$ for BS\,801. Although they are not as rapid
rotators as $\delta$ Sco (v\,$\sin i$ = 148--180 km\,s$^{-1}$ in the
same catalog), we used these stars because their OHP spectra are
clean and contain no other photospheric lines in the region between
4450 and 4500 \AA\ except for the He {\sc i} 4471 \AA\ and Mg {\sc
ii} 4482 \AA. Intensity ratio of these two lines is one of the main
temperature criteria for B--type stars. The Mg {\sc ii} line
strengthens and the He {\sc i} line weakens as the photospheric
temperature decreases.

The composite spectrum may not be an accurate representation of the
real situation due to uncertainties of the secondary component
properties, but a broadening of the He {\sc i} 4471\,\AA\ line
profile was detected on the periastron day (Fig.\,\ref{f3}b). This
result can be interpreted as a detection of the secondary, although
its parameters cannot be further constrained. However, it does not
contradict the current view of this component \citep[see
also][]{Meilland2011,Che2012}.

\subsection{The binary components masses and the system
distance}\label{masses}

There are several estimates for the mass of the primary component of
$\delta$ Sco (M$_1$) based on recent models of stellar evolution for
the solar metallicity \citep[e.g.,][]{Ekstrom2012} with \citep[13.9
M$_\odot$,][]{Che2012} or without \citep[14.6
M$_\odot$,][]{Pecaut2012} taking into account gravity darkening due
to the fast rotation of the star. Combining the spectroscopic
orbital solution derived here with the interferometry results
\citep{Che2012}, one can estimate the components masses separately
and derive the distance toward the system.

In order to do this, we calculate the system mass function
(0.244$\pm$0.025 M$_\odot$) from the data presented in Table
\ref{t3}. We also take into account the contribution of the
secondary to the observed flux (see Sect. \ref{intro}), which was
mentioned but not considered by \citet{Pecaut2012}. This reduces the
luminosity of the primary by 0.08 dex and its mass by 0.2
M$_{\odot}$ to M$_1$ = 14.4 M$_{\odot}$ in the non-rotating case.

The components mass ratio ($q_{1,2}$) depends on the orbital
inclination angle ($i$) and the mass function. It requires $i =
36\degr \pm 1\degr$ to be consistent with the brightness ratio and
comes to $q_{1,2} = 1.67\pm0.07$, giving an estimate for the
secondary mass of M$_{2} = 8.6\pm0.6$ M$_{\odot}$. The total mass of
the binary implies an orbital semi-major axis of 13.8$\pm$0.2 AU The
latter combined with the angular measure of the semi-major axis from
interferometry \citep[99.04 mas,][]{Che2012} gives a distance of
140.0$\pm$1.5 pc toward the system.

Rapid rotation creates a temperature distribution on a star's
surface that changes its fundamental parameters depending on the
rotation rate and viewing angle \citep[gravity
darkening,][]{Townsend2004}. For $\delta$ Sco that is viewed at an
intermediate angle and rotates at $\sim$60\% of the critical rate
\citep[estimated from v\,$\sin i$ = 148 km\,s$^{-1}$,][]{Brown1997},
this effect increases its apparent luminosity by $\sim$0.2 dex and
barely affects its apparent color. Taking this into account reduces
M$_1$ to 13 M$_\odot$ and for the same $i = 36\degr \pm 1\degr$
changes $q_{1,2}$ to 1.6$\pm$0.1, M$_2$ to 8.2$\pm$0.6 M$_\odot$,
the orbital semi-major axis to 13.5$\pm$0.1 AU, and the distance to
136.0$\pm$1.5 pc.

Our estimates for M$_2$ are consistent with its adopted spectral
type (B1--B3, T$_{\rm eff}$ = 20000--24000 K), while the distance
toward the system is consistent with the HIPPARCOS parallax values
from both published solutions \citep[123$^{+15}_{-12}$ and
151$^{+23}_{-18}$ pc,][respectively]{ESA1997,vanLeeuwen2007}. The
new distance estimates are accounted for in the components
luminosity calculation. This procedure resulted in a slightly larger
orbital inclination angle compared to those determined from
interferometry \citep[$30\fdg2\pm0\fdg7$, $32\fdg9\pm0\fdg2$,
$32\fdg3\pm0\fdg3$,]
[respectively]{Meilland2011,Tycner2011,Che2012}. If $i \le$ 33\degr,
then M$_{2} \ge$ 10 M$_{\odot}$ that is inconsistent with the
reported components brightness ratio. No gravity darkening
correction was applied to the secondary fundamental parameters,
because no information about its rotation is currently available.

\begin{figure}
\includegraphics[width=8cm,height=7.2cm]{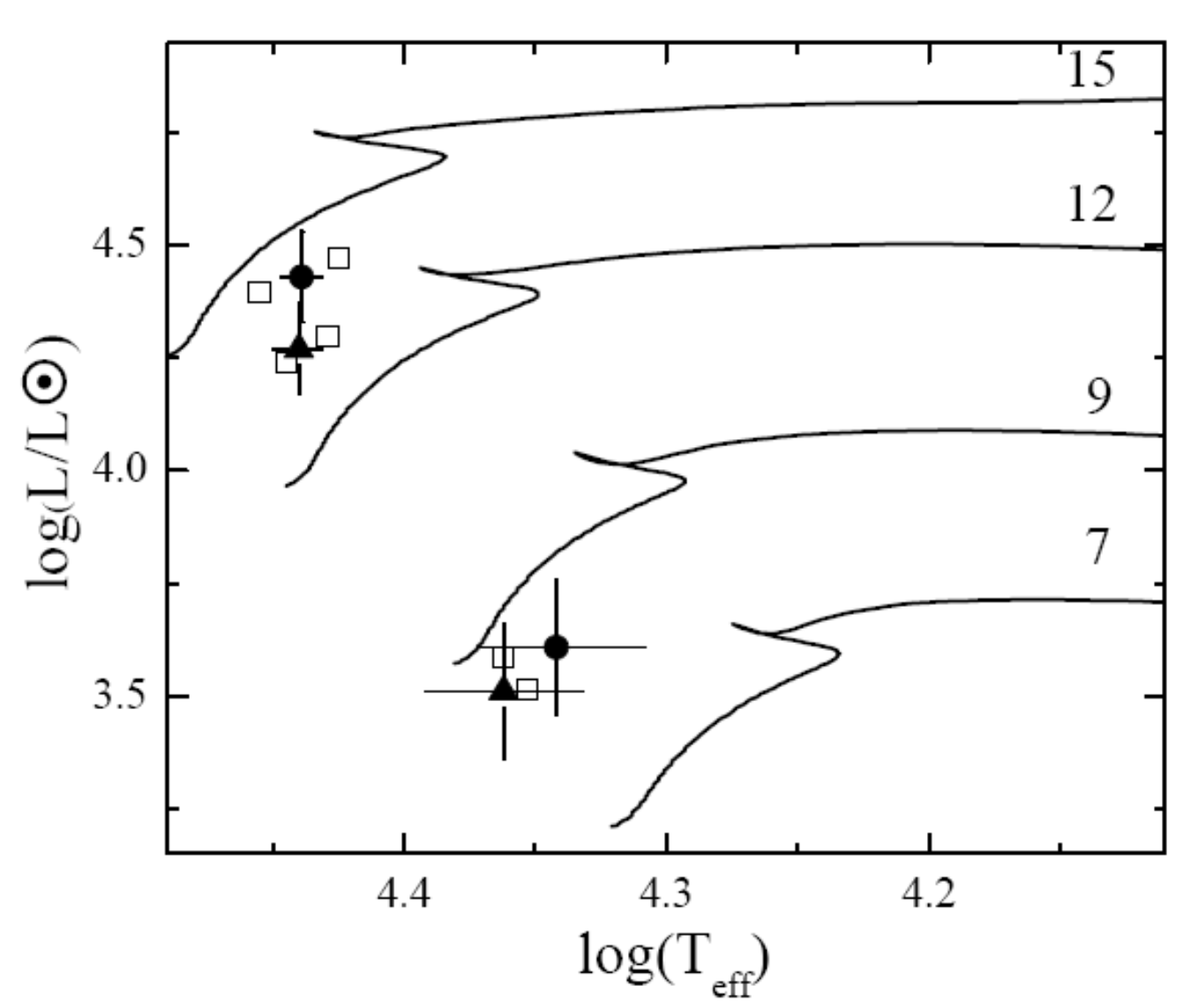}
\caption{Part of a Hertzsprung--Russell diagram showing evolutionary
tracks with rotation from \citet{Ekstrom2012}. Initial masses in
solar units are indicated by the numbers near corresponding tracks.
No gravity darkening positions of the $\delta$ Sco A and B
components are shown by filled circles, while those with this effect
accounted are shown by filled triangles. Open squares near component
A positions indicate theoretical values for ages of 8 and 10 Myrs.
Open squares near component B positions indicate theoretical values
with an age of 10 Myrs for initial masses of 8.6 M$_{\odot}$ (upper
square) and 8.2 M$_{\odot}$ (lower square). \label{f4}}
\end{figure}

Also, the above fundamental parameters imply an age of 9--10 Myrs
for both rotating and non-rotating primary mass estimates
(Fig.\,\ref{f4}). The theoretical positions for both secondary
component models (8.6 M$_\odot$ and 8.2 M$_\odot$) with an age of 10
Myrs are within the uncertainties of our adopted parameters. The
discrepancy does not seem to be dramatic and may be due to the
unknown mass transfer history for the $\delta$ Sco system as well as
not well constrained spectral properties of the secondary component.

Both sets of the mass estimates are close to each other, but the one
with gravity darkening taken into account seems to be more
realistic. This subject will be explored in more detail in a
separate study of the system's spectral energy distribution.


\section{A New View of the $\delta$ Sco System}
\label{discussion}

There are several facts that make $\delta$ Sco unusual among Be
binaries. First, the Bright Star Catalog \citep{Hoffleit1991}
mentions a possible companion with a 20\,day orbital period in the
system \citep[initially reported by][]{vanHoof1963}. Second, most Be
binaries with a non-degenerate secondary component have circular
orbits \citep[e.g.,][]{Bjorkman2002,Nemravova2012}, while most
Be/X-ray binaries have eccentric orbits \citep[e.g.,][]{Reig1997}.
Third, radial velocities of $\delta$ Sco reported in several papers
throughout the 20th century show variations additional to those
expected at periastra (see Sect. \ref{historic}). The difference
between the 2000 and 2011 radial velocity curves reported above may
be part of the same issue. Finally, thanks to the {\it IRAS} survey
of bow shocks around runaway stars \citep{vanBuren1995}, $\delta$
Sco is known as a bow-shock-producing star.

Taken together, the above facts allow us to suggest the hypothesis
that $\delta$ Sco might be a runaway triple system
\citep[cf.,][]{Gvaramadze2012}. A number of Be stars are known to be
part of triple and more complex stellar systems
\citep[e.g.,][]{Koubsky2010}. Below we discuss the above mentioned
facts more closely and conclude whether our hypothesis is viable.

\subsection{Possible origin of the high orbital eccentricity and the runaway status
of the $\delta$ Sco system} \label{reasons}

The system orbital eccentricity is so large that component A (the Be
star) and B (the interferometrically detected one) at periastron
come together at a distance of $\le$ 0.8 AU There are several
processes that can lead to such a highly eccentric orbit: a) close
dynamical encounter between a binary system and another binary or a
single star \citep[e.g.,][]{Hills1975,Hoffer1983}, b) dynamical
decay of a trapezium-like system \citep{Allen1974}, and c) the
Lidov-Kozai resonance in a hierarchical triple system
\citep{Lidov1962,Kozai1962}.

The first process could also be responsible for ejection of $\delta$
Sco from the parent star cluster
\citep[e.g.,][]{Leonard1990,Kroupa1998}. The ejection event might
have occurred when the cluster was much smaller in size and much
denser. This scenario is consistent with the presence of the bow
shock-like structure around $\delta$ Sco which is likely the result
of the system (super)sonic motion through the interstellar medium.

To check the runaway status of $\delta$ Sco, we used the proper
motion measurements for this system from the new reduction of the
{\it Hipparcos} data by \citet{vanLeeuwen2007}, $\mu _\alpha \cos
\delta = -10.21\pm 1.01$ mas ${\rm yr}^{-1}$, $\mu _\delta =
-35.41\pm 0.71$ mas ${\rm yr}^{-1}$ or, in Galactic coordinates,
$\mu _l = -32.22\pm0.87$ mas ${\rm yr}^{-1}$ and $\mu _b = -17.83\pm
0.87$ mas ${\rm yr}^{-1}$. After correction for the Galactic
differential rotation and the solar peculiar motion\footnote{Here we
used the Galactic constants $R_0$=8.0 kpc and $\Theta _0$=$240 \,
{\rm km} \, {\rm s}^{-1}$ \citep{Reid2009} and the solar peculiar
motion $(U_{\odot},V_{\odot},W_{\odot})=(11.1,12.2,7.3) \, {\rm km}
\, {\rm s}^{-1}$ \citep{Schonrich2010}.}, this proper motion
translates into the transverse peculiar velocity $v_{\rm tr} = (v_l
^2 +v_b ^2 )^{1/2} = 11.1\pm0.6 \, {\rm km} \, {\rm s}^{-1}$, where
$v_l = -7.0\pm0.6 \, {\rm km} \, {\rm s}^{-1}$ and $v_b = -8.6\pm0.6
\, {\rm km} \, {\rm s}^{-1}$. To this velocity one should add a
peculiar radial velocity of $4.9\pm0.2 \, {\rm km} \, {\rm s}^{-1}$,
derived from the systemic radial velocity of $-6.7\pm0.2 \, {\rm km}
\, {\rm s}^{-1}$ \citep{Meilland2011}, so that the total
(three-dimensional) velocity of $\delta$ Sco $v_* \approx 12 \, {\rm
km} \, {\rm s}^{-1}$, and its vector is inclined to the plane of the
sky by an angle of $\approx 34\degr$. Thus, $\delta$ Sco is a
low-velocity runaway system \citep[cf.][]{Gvaramadze2012a}. In their
Figure 4, \citet{Peri2012} presented a 22\,$\mu$m image of an
arc-like structure around $\delta$ Sco taken by the {\it Wide-field
Infrared Survey Explorer} \citep[{\it WISE},][]{Wright2010}. One can
see that the vector of the peculiar transverse velocity of the star
is almost parallel to the symmetry axis of the structure, which
supports its interpretation as a bow shock.

$\delta$ Sco is located within the confines of the Sco\,OB2
association, and the distances to both systems are comparable to
each other. This might imply that Sco\,OB2 is the parent association
of $\delta$ Sco. To check this possibility, one should compare the
proper motion of $\delta$ Sco with that of Sco\,OB2, $\mu _l =
-23.5\pm 2.9$ mas ${\rm yr}^{-1}$ and $\mu _b = -10.5\pm 4.0$ mas
${\rm yr}^{-1}$ \citep{Melnik2009}, which also is based on the new
reduction of the {\it Hipparcos} data by \citet{vanLeeuwen2007}.
This comparison shows that although both objects are moving almost
in the same direction, there is a significant difference in the
magnitudes of their proper motions. This difference could partially
be caused by the effect of binarity of $\delta$ Sco, because, as
noted by \citet{deZeeuw1999}, ``the {\it HIPPARCOS} proper motion,
observed during the mission lifetime of $\sim 3.3$ yr, does not
necessarily reflect the center-of-mass proper motion''. This
difference, however, could also be caused by the runaway status of
$\delta$ Sco. Moreover, one cannot exclude the possibility that the
effect of binarity could instead be responsible for some {\it
reduction} of the difference in the proper motions. Taken at face
value, the observed proper motions imply that $\delta$ Sco is moving
with a transverse velocity of $7.5 \, {\rm km} \, {\rm s}^{-1}$ with
respect to the association. From this, in turn, it follows that the
birth place of this $\sim 10$ Myr old star should be at $\approx
30\degr$ from the birth place of the association (provided that the
age of the association is $\sim 10$ Myr as well). We conclude
therefore that it is likely that $\delta$ Sco was injected in the
Sco\,OB2 association from outside
\citep[cf.][]{Gvaramadze2011,Gvaramadze2012a} and that the parent
cluster of this star is located at $\sim 50\degr$ from its current
position.

We also note that it has recently been found that the orbital
momentum of the secondary and the disk's rotation vector are
opposite to each other \citep{Stefl2012,Che2012}. This might imply
that the $\delta$ Sco system was formed in the course of a few-body
dynamical interaction or could be a consequence of the Lidov-Kozai
resonance, if the system is triple. Detailed discussion of these
possibilities, however, is beyond the scope of this paper.

\subsection{Stability of the binary orbital period}\label{historic}

\begin{figure}
\includegraphics[width=15cm,height=8cm]{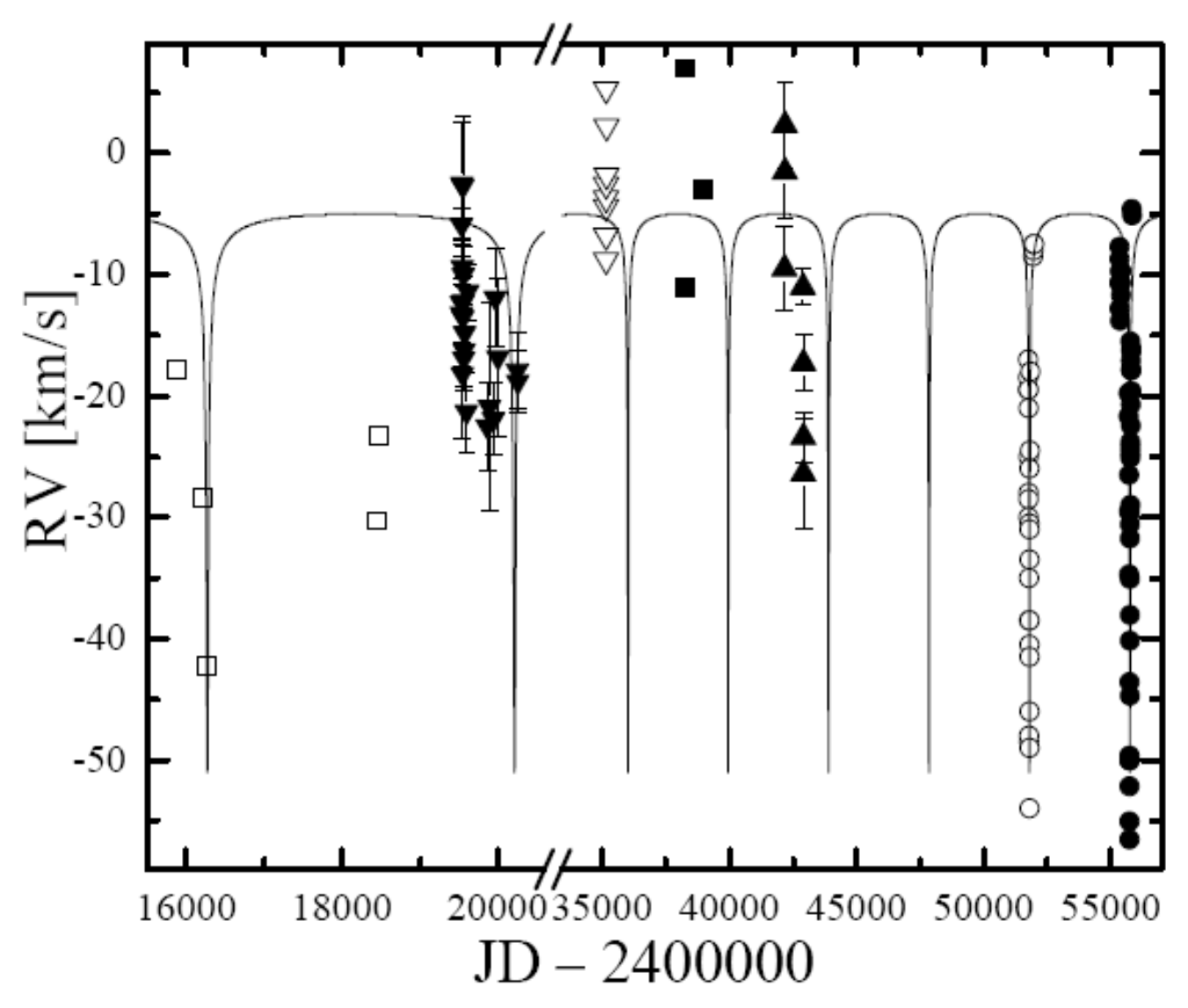} \caption{Radial
velocities of $\delta$ Sco measured since the early 20--th century.
Symbols: open squares are data from \citet{Frost1926}, filled
downward triangles -- \citet{Beardsley1969}, open downward triangles
-- \citet{vanHoof1963}, filled squares -- \citet{Thackeray1966},
filled upward triangles -- \citet{Levato1987}, open circles --
\citet{Miroshnichenko2001}, and filled circles -- this work. The
solid line represents the orbital solution for the 2011 periastron
shown in Table \ref{t3} (see Sect. \ref{rv_curve}). \label{f5}}
\end{figure}

We will use radial velocity data published during the 20th century
\citep{Frost1926,vanHoof1963,Thackeray1966,Beardsley1969,Levato1987}
to search for the binary orbital period variations. These papers
present 49 measurements obtained between 1902 and 1976. The data
along with those from \citet{Miroshnichenko2001} and this work are
shown in Fig. \ref{f5}. Most of the historical data deviate from the
expected radial velocity behavior based on the modern solution. This
can be partly due to both random and systematic errors. For example,
\citet{Beardsley1969} concluded that the Allegheny observatory data
presented in his work showed no variations and can be averaged.
Still, some data points from \citet{Frost1926} and
\citet{Beardsley1969} match with the expected radial velocity minima
due to the periastron passages.

We investigated both the historical (20th century only) and full
data sets using the methods for finding periodicity from
\citet{Dworetsky1983} and \citet{Scargle1982}. Both methods are
suited for unevenly spaced data, but Scargle's method works better
for nearly sinusoidal variations. The historical data miss most of
the radial velocity minima that result in not well-defined extrema
of the periodograms. The Dworetsky ``string-length'' function shows
the deepest minimum for a period of 3750 days for the historical
data, while the Scargle periodogram shows the highest maximum for a
period of 3339 days. In both cases, it is at least 200 days shorter
than the one determined from the periastra of 2000 and 2011.

Nevertheless, both methods detected the modern period when fitting
the entire data collection. Dworetsky's function peaks at 3928 days,
while Scargle's one at 3915 days. The shifts toward shorter periods
are due to the historical data. They can be caused by the variable
orbital period and/or errors in the radial velocity measurements.

Although the existing material does not allow us to make a definite
conclusion about the reality of these shifts, it prompts us to
analyze the role of a possible third component in the system in the
origin of changes in the orbital period. We also examine the
stability of triple systems and show that under the certain
conditions the presence of the third component could be responsible
for maintaining the system's high eccentricity.

\subsection{Analysis of the long-term radial velocity variations}
\label{triple}

Let us now consider stability of triple systems that may result in
the observed radial velocity variations. Orbital solutions in
general require numerical integration, but approximations can be
obtained in some cases. In particular, stability conditions for
hierarchical triple systems with a low eccentricity of the inner
pair are given by the following formula \citep{Valtonen2008}:

\begin{equation}\label{eq1}
    \left(\frac {q} {a_i}\right) >3\left(1+\frac{m_3}{m_1+m_2}\right)^{1/3}(1-e)^{-1/6}\left(\frac{1+\cos \iota }
    2\right)^{1/3}
\end{equation}

where $m_1$ and $m_2$ are component masses of the main binary; $a_i$
is the semi-major axis of the main binary; $m_3$ is the mass of the
third component; $e$ is the orbital eccentricity of the third
component; $q$ is the periastron distance of the third component
orbit from the barycenter of the main binary; and $i$ is the
relative inclination of the third component orbit. This result can
be used to investigate whether the component A is a binary system
with an orbital period of 20 days \citep[as was suggested
by][]{vanHoof1963}.

In other words, we consider a hierarchical system with an inner
binary A--C and an external component B. Assuming $m_{1} = 13$
M${_\odot}$ (see Sect. \ref{masses}), the semi-major axis of a body
at a $T$ = 20--day orbit is $\sqrt[3]{Gm_1T^2/(4\pi ^2)} = 0.35$ AU
Such a triple system is stable if component B does not come closer
than 1.65 AU to the center of the A--C pair. The observations show
that components A and B get as close together as 0.8 AU
\citep[e.g.,][]{Miroshnichenko2001,Che2012}. The ultimate fate of
such a system is a breakdown at already the second revolution. We
therefore concur with \citet{Miroshnichenko2001} that the 20 day
period is spurious.

Next we modeled the evolution of a stable triple system with a
component C that is external to the eccentric A--B binary by
numerically solving a system of motion equations in Cartesian
coordinates. We used the orbital parameters of the inner binary from
Table \ref{t3}. The masses were assumed to be 13 and 8.2 M$_\odot$
for the components A and B, respectively (see Sect. \ref{masses}).
The goal was to constrain parameters of component C, which can be
responsible for the observed difference of the radial velocity
curves obtained in 2011 and in 2000 as well as for long-term period
variations. The results show that there is no unique solution to
this problem, although certain constraints on the external orbit can
be placed. In particular, the semi-major axis of the component C
orbit cannot be smaller than 67.5 AU and, therefore, its orbital
period should be longer than 120 years.

\begin{figure}
\begin{tabular}{cc}
\includegraphics[width=7.3cm,height=5cm]{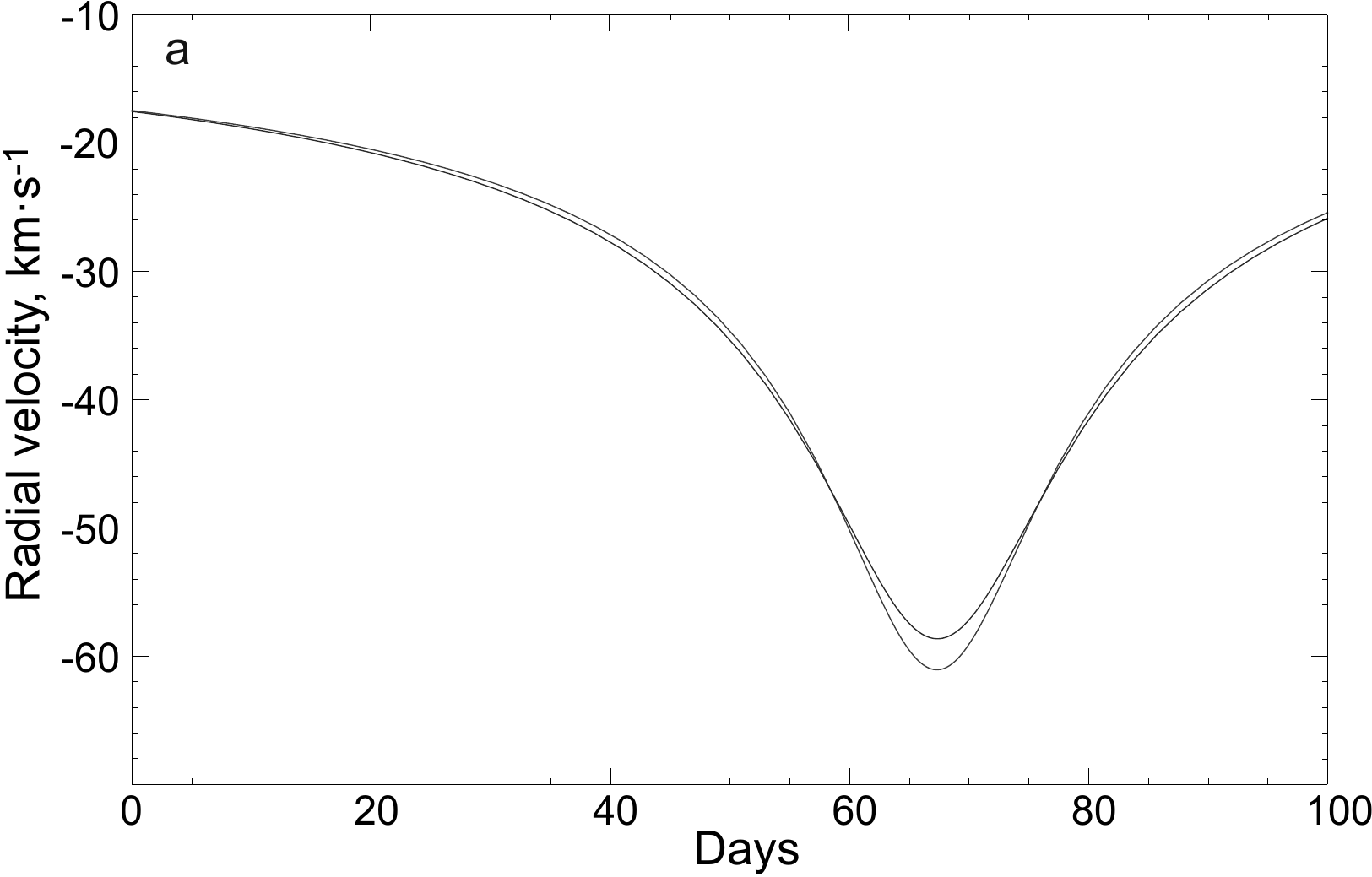}&
\includegraphics[width=8.3cm,height=5cm]{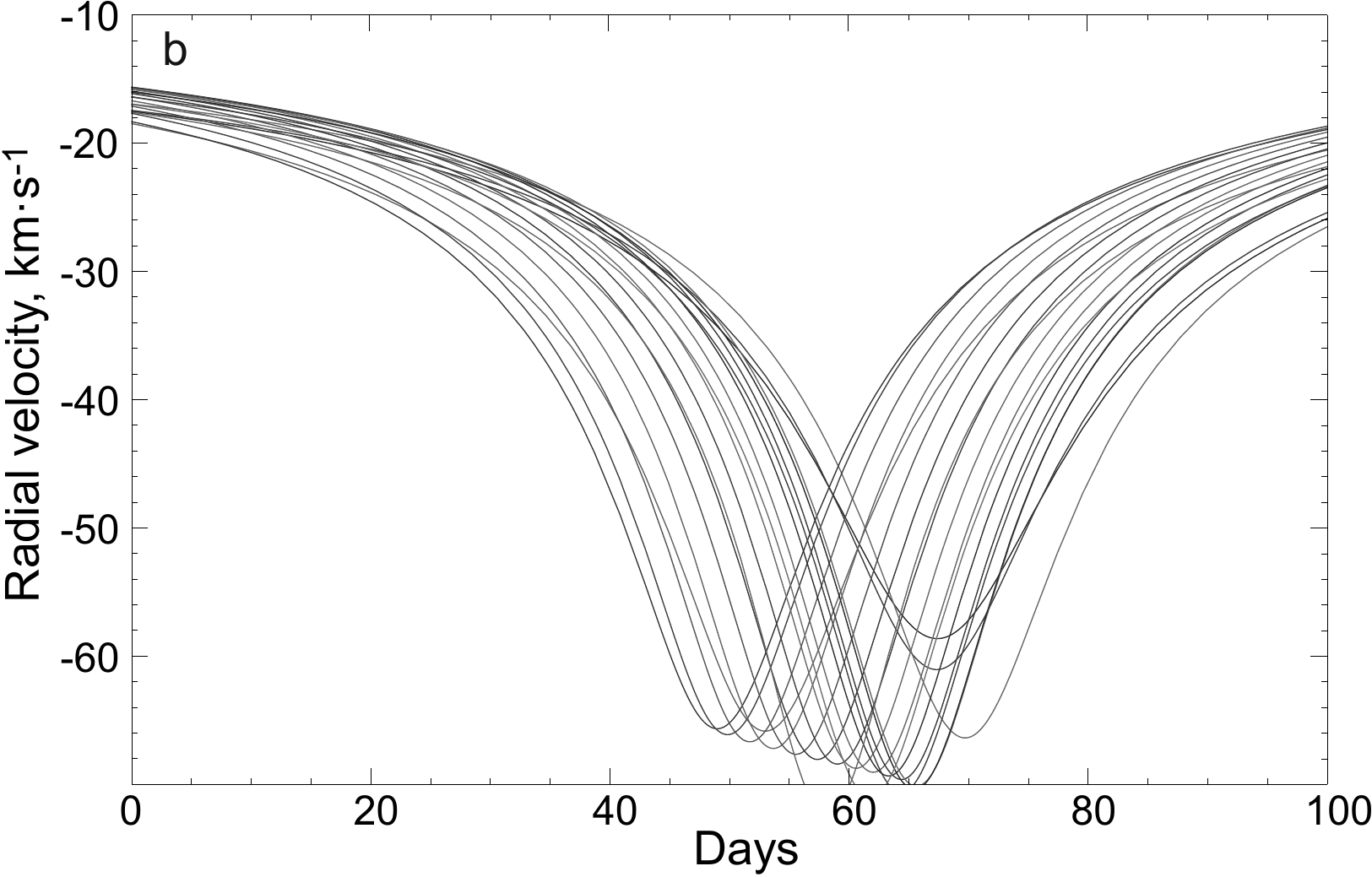}\\
\end{tabular}
\caption{Panel a) shows radial velocity curves for two consecutive
orbital cycles of a triple system that consists from an inner binary
and an outer third component with the orbital parameters described
in Sect. \ref{triple}. Panel b) shows radial velocity curves for
twenty consecutive orbital cycles of the inner binary from the same
triple system that occur during one orbital period of the outer
component. Time scale has an arbitrary zero-point. \label{f6}}
\end{figure}

As an example, we calculated radial velocity curves for a described
above triple system with an external component C that has the
following parameters: mass M$_{\rm C}$ = 1.5 M$_\odot$, semi-major
axis $a = 100$ AU, eccentricity $e = 0.5$, inclination of the orbit
to the orbital plane of the inner binary $i_{\rm C} = 60\degr$,
periastron longitude $\omega_{\rm C} = 30\degr$, and the mean
anomaly $E = 0\degr$ for 2011 July 4. Two consecutive orbital cycles
of this system demonstrate that the radial velocity curve can change
similarly to the observed behavior for $\delta$ Sco (Fig.
\ref{f6}a), while twenty of them show how orbital period can change
over one orbit of the external component(Fig. \ref{f6}b).

The mechanism responsible for the orbital variations of the inner
binary is known as the above mentioned Lidov-Kozai resonance. It
occurs in hierarchical triple systems, in which the external
component orbit is tilted with respect to that of the inner binary.
This situation leads to precession of the orbits around the
direction of the total moment of impulse of the system.
Additionally, if the inclination between the inner and the outer
orbit exceeds $\sqrt{\arccos (3/5)}$, the precession is accompanied
by the angular momentum exchange between the orbits. The latter
manifests itself by cyclic variations of both orbit eccentricities.
A detailed analysis of this effect is given by \citet{Valtonen2006}.

Since the evolution of a binary system due to tidal friction could
lead to a secular decrease of the orbital eccentricity
\citep[e.g.,][]{Zahn2008}, it is likely that the large current
eccentricity of the $\delta$ Sco system implies an inclination of
the component C orbit of $\ge 40\degr$. This limits the set of
orbital solutions capable of explaining the observed radial velocity
variations.

Component C can be located as far as 0\farcs3 to 0\farcs8 from the
center of mass of the inner binary, but it has not been detected
yet. We can suggest several explanations for this result. First, the
component C orbital period is very long (213 years for the orbital
solution shown in Fig. \ref{f6}). If component C is a normal star
born in the same cluster, it should be nearly 8 mag fainter
\citep[for M = 1.5 M$_\odot$,][]{Ekstrom2012} than component A. Such
a faint star can probably be directly detected by interferometry,
which has only been used to observe the system for the last $\sim$
40 years. During this time period, component C might have been
projectionally close to the inner binary. Also, the high brightness
contrast might have hampered the detection.

\section{Conclusions} \label{conclusions}

The spectroscopic observational campaign of the 2011 periastron
passage in the $\delta$ Sco system was successful and resulted in a
new measurement of the orbital period ($10.8092\pm 0.0005$ years or
$3948.0\pm1.8$ days). We also determined orbital elements using the
radial velocity curve for the He {\sc ii} 4686 \AA\ line which
turned out to be very close to those found from the H$\alpha$ radial
velocity data during the previous periastron passage in 2000. Using
the new orbital solution and recent evolutionary models with
rotation \citep{Ekstrom2012}, we derived new masses of the binary
components corrected for gravity darkening of the primary (13 and
8.2 M$_\odot$). These estimates along with the refined angular
semi-major axis of the binary orbit \citep{Che2012} imply that
$\delta$ Sco is located at a distance of 137 pc from the Sun that is
consistent with both alternative solutions for the HIPPARCOS
parallax. The radial velocity and line profile variations observed
in the H$\alpha$ line near the 2011 periastron were affected by the
interaction of the secondary component and the CS disk around the
primary component. Using these data, we estimated a disk radius of
150 R$_\odot$. This result is consistent with previous
interferometric measurements of the disk radius
\citep[e.g.,][]{Meilland2011} and indicates that the disk was most
likely growing during the entire time between the last two
periastron passages.

We have detected a signature of the secondary component, but not
clearly revealed its properties. This result is consistent with an
early B--type spectral type for this component. The radial velocity
curve for the He {\sc ii} line slightly deviates from that derived
in 2000 for the H$\alpha$ line. This result along with the high
eccentricity of the system and the presence of a bow shock-like
structure around it suggest that $\delta$\,Sco might have a third
component, external to the interferometric binary, and be a runaway
object, dynamically ejected from its parent cluster. If the third
component is indeed present, then the orbital elements presented in
Table \ref{t3} may change in the future. The system needs constant
observational attention to verify this suggestion.

Finally, the campaign confirmed that amateur spectroscopy becomes an
important factor in astronomy of emission-line stars. It is very
important to continue observing $\delta$\,Sco spectroscopically,
photometrically, and interferometrically to search for more clues
about the nature of this unusual stellar system.

\acknowledgements This work was supported in part by DGAPA/PAPIIT
project IN 103912. A.~M. acknowledges financial support from the
University of North Carolina at Greensboro and from its Department
of Physics and Astronomy. A.~M., J.~R., A.~F., and T.~G. thank the
Instituto de Astro\'\i sica de Canarias and its staff for allocating
observing time and for technical, logistical and financial support,
as well as for their hospitality during the observing run at the
Teide Observatory. J.~R. thanks Alex Abad, Emilio Cadavid and all
the staff that they direct for the technical support in preparation
and during the mission at the IAC80 telescope. A.~C.~C. acknowledges
support from CNPq (grant 308985/2009--5) and Fapesp (grant
2010/19029--0). This research has made use of the SIMBAD database,
operated at CDS, Strasbourg, France and the BeSS database, operated
at LESIA, Observatoire de Meudon, France (accessible at
http://basebe.obspm.fr).

\begin{table}
\caption[]{Radial velocities for the H$\alpha$ line} \label{t1}
\begin{center}
\begin{tabular}{rcrlc}
\hline\noalign{\smallskip}
Date        & JD2450000+  &  RV        & Error      & Source\\
MM/DD/YYYY  & days        &km\,s$^{-1}$&km\,s$^{-1}$&       \\
\noalign{\smallskip}\hline\noalign{\smallskip}
05/28/2010 &5344.835    &   -2.6    &   0.5     &   1   \\
05/30/2010 &5346.835    &   -1.7    &   0.6     &   1   \\
06/03/2010 &5350.779    &   -2.0    &   0.5     &   1   \\
06/04/2010 &5351.802    &   -2.5    &   0.2     &   1   \\
06/19/2010 &5366.766    &   -3.5    &   0.1     &   1   \\
06/22/2010 &5369.747    &   -2.7    &   0.3     &   1   \\
06/25/2010 &5372.901    &   -4.0    &   0.1     &   1   \\
07/22/2010 &5399.741    &   -0.5    &   0.3     &   1   \\
07/25/2010 &5402.743    &   -1.3    &   0.2     &   1   \\
07/27/2010 &5404.763    &   -0.6    &   0.1     &   1   \\
07/31/2010 &5408.742    &   -1.2    &   0.2     &   1   \\
08/05/2010 &5413.842    &   -2.3    &   0.9     &   1   \\
10/05/2010 &5474.929    &   -3.5    &   0.6     &   6   \\
10/10/2010 &5479.927    &   -3.8    &   1.5     &   6   \\
10/21/2010 &5490.913    &   -6.0    &   0.3     &   6   \\
01/10/2011 &5571.762    &   -15.1   &   3.4     &   4   \\
01/16/2011 &5577.729    &   -14.5   &   3.0     &   4   \\
01/20/2011 &5581.745    &   -13.4   &   3.1     &   4   \\
01/23/2011 &5584.749    &   -14.1   &   3.0     &   4   \\
01/23/2011 &5585.363    &   -14.9   &   2.2     &   2   \\
01/28/2011 &5589.704    &   -15.7   &   1.6     &   8   \\
01/30/2011 &5591.707    &   -12.8   &   1.8     &   8   \\
02/09/2011 &5601.725    &   -11.2   &   3.0     &   4   \\
03/07/2011 &5628.349    &   -6.7    &   2.3     &   2   \\
03/09/2011 &5629.641    &   -14.1   &   2.7     &   9   \\
03/10/2011 &5630.658    &   -14.1   &   3.0     &   9   \\
03/13/2011 &5634.399    &   -7.3    &   2.2     &   2   \\
03/16/2011 &5637.392    &   -7.4    &   2.8     &   2   \\
03/29/2011 &5649.566    &   -17.7   &   1.2     &   8   \\
04/05/2011 &5656.605    &   -17.5   &   2.4     &   9   \\
04/06/2011 &5657.558    &   -19.1   &   2.6     &   9   \\
04/09/2011 &5660.542    &   -13.9   &   4.7     &   5   \\
04/15/2011 &5667.488    &   -11.0   &   3.6     &   5   \\
\noalign{\smallskip}\hline \noalign{\smallskip}
\end{tabular}
\end{center}
\end{table}

\clearpage \noindent {\bf Table 1. continued}
\begin{center}
\begin{tabular}{rcrlc}
\hline\noalign{\smallskip}
Date        & JD2450000+    &  RV        & Error    & Source\\
MM/DD/YYYY  & days          &km\,s$^{-1}$&          &       \\
\noalign{\smallskip}\hline \noalign{\smallskip}
04/16/2011  &5668.391    &   -6.4    &   1.1     &   2   \\
04/25/2011  &5677.378    &   -7.0    &   1.1     &   2   \\
04/27/2011  &5678.600    &   -13.2   &   3.9     &   5   \\
04/30/2011  &5681.561    &   -16.5   &   5.6     &   5   \\
05/03/2011  &5685.514    &   -11.1   &   4.9     &   4   \\
05/11/2011  &5692.535    &   -11.7   &   5.2     &   4   \\
05/20/2011  &5702.463    &   -10.9   &   1.0     &   5   \\
05/22/2011  &5704.443    &   -16.8   &   5.1     &   4   \\
05/27/2011  &5709.507    &   -13.0   &   0.9     &   5   \\
05/28/2011  &5709.573    &   -12.1   &   0.4     &   4   \\
05/29/2011  &5710.400    &   -11.9   &   3.3     &   4   \\
06/01/2011  &5713.960    &   -14.5   &   0.8     &   6   \\
06/01/2011  &5714.371    &   -23.5   &   2.4     &   8   \\
06/02/2011  &5715.381    &   -15.8   &   1.6     &   8   \\
06/08/2011  &5720.529    &   -18.3   &   2.0     &   1   \\
06/08/2011  &5721.026    &   -18.4   &   2.1     &   1   \\
06/08/2011  &5721.395    &   -24.7   &   2.0     &   8   \\
06/09/2011  &5722.407    &   -19.7   &   1.8     &   1   \\
06/11/2011  &5723.797    &   -19.1   &   0.3     &   6   \\
06/11/2011  &5724.445    &   -22.8   &   3.9     &   7   \\
06/12/2011  &5725.318    &   -21.6   &   1.9     &   1   \\
06/14/2011  &5727.387    &   -32.3   &   2.4     &   8   \\
06/14/2011  &5727.436    &   -29.6   &   4.8     &   4   \\
06/14/2011  &5727.446    &   -26.7   &   2.6     &   5   \\
06/15/2011  &5728.459    &   -33.7   &   7.3     &   4   \\
06/16/2011  &5729.269    &   -26.9   &   1.8     &   1   \\
06/16/2011  &5729.380    &   -43.0   &   3.2     &   8   \\
06/18/2011  &5731.406    &   -31.2   &   1.3     &   5   \\
06/21/2011  &5734.412    &   -42.5   &   3.6     &   8   \\
06/22/2011  &5735.087    &   -34.9   &   1.8     &   2   \\
06/22/2011  &5735.380    &   -42.2   &   3.9     &   8   \\
06/24/2011  &5737.394    &   -45.4   &   4.5     &   8   \\
06/26/2011  &5738.984    &   -41.9   &   1.2     &   2   \\
\noalign{\smallskip}\hline \noalign{\smallskip}
\end{tabular}
\end{center}

\clearpage \noindent {\bf Table 1. continued}

\begin{center}
\begin{tabular}{rcrlc}
\hline\noalign{\smallskip}
Date        & JD2450000+    &  RV        & Error    & Source\\
MM/DD/YYYY  & days          &km\,s$^{-1}$&          &       \\
\noalign{\smallskip}\hline \noalign{\smallskip}
06/26/2011  &5739.387    &   -49.4   &   0.6     &   8   \\
06/27/2011  &5740.374    &   -52.2   &   0.3     &   8   \\
06/28/2011  &5741.503    &   -49.4   &   1.0     &   3   \\
06/29/2011  &5741.939    &   -51.0   &   0.4     &   6   \\
06/29/2011  &5742.396    &   -49.5   &   0.8     &   3   \\
07/01/2011  &5743.905    &   -50.6   &   1.5     &   6   \\
07/01/2011  &5744.301    &   -49.4   &   0.7     &   1   \\
07/02/2011  &5744.555    &   -49.2   &   0.3     &   3   \\
07/02/2011  &5745.241    &   -47.9   &   0.9     &   1   \\
07/03/2011  &5745.502    &   -48.1   &   0.3     &   4   \\
07/03/2011  &5745.543    &   -47.4   &   0.5     &   3   \\
07/03/2011  &5746.119    &   -47.4   &   0.6     &   6   \\
07/03/2011  &5746.241    &   -48.4   &   1.4     &   1   \\
07/04/2011  &5746.521    &   -47.0   &   1.4     &   3   \\
07/04/2011  &5746.938    &   -50.0   &   1.5     &   6   \\
07/04/2011  &5747.377    &   -49.4   &   2.3     &   4   \\
07/04/2011  &5747.428    &   -49.2   &   1.9     &   1   \\
07/05/2011  &5747.549    &   -48.8   &   1.6     &   3   \\
07/05/2011  &5748.383    &   -52.6   &   1.1     &   8   \\
07/05/2011  &5748.407    &   -47.9   &   2.1     &   1   \\
07/06/2011  &5748.548    &   -48.6   &   1.7     &   3   \\
07/06/2011  &5749.122    &   -49.2   &   2.0     &   2   \\
07/07/2011  &5749.549    &   -47.6   &   1.5     &   3   \\
07/07/2011  &5750.409    &   -44.5   &   0.2     &   8   \\
07/07/2011  &5750.432    &   -49.0   &   0.8     &   1   \\
07/08/2011  &5750.546    &   -46.7   &   1.8     &   3   \\
07/08/2011  &5751.372    &   -44.7   &   1.6     &   8   \\
07/09/2011  &5751.880    &   -41.5   &   1.0     &   6   \\
07/09/2011  &5752.259    &   -39.6   &   0.5     &   1   \\
07/10/2011  &5753.090    &   -42.0   &   1.0     &   2   \\
07/11/2011  &5754.085    &   -44.8   &   1.5     &   2   \\
07/11/2011  &5754.366    &   -50.1   &   3.4     &   8   \\
07/13/2011  &5755.976    &   -44.1   &   1.3     &   2   \\
\noalign{\smallskip}\hline \noalign{\smallskip}
\end{tabular}
\end{center}

\clearpage \noindent {\bf Table 1. continued}

\begin{center}
\begin{tabular}{rcrlc}
\hline\noalign{\smallskip}
Date        & JD2450000+    &  RV        & Error    & Source\\
MM/DD/YYYY  & days          &km\,s$^{-1}$&          &       \\
\noalign{\smallskip}\hline \noalign{\smallskip}
07/13/2011  &5756.189    &   -43.9   &   1.1     &   2   \\
07/13/2011  &5756.281    &   -44.2   &   1.3     &   1   \\
07/14/2011  &5756.897    &   -42.1   &   1.0     &   6   \\
07/14/2011  &5757.211    &   -43.6   &   0.7     &   2   \\
07/14/2011  &5757.388    &   -43.6   &   1.0     &   1   \\
07/15/2011  &5757.882    &   -46.5   &   2.4     &   6   \\
07/15/2011  &5758.371    &   -45.9   &   1.2     &   8   \\
07/17/2011  &5759.961    &   -38.3   &   1.0     &   2   \\
07/17/2011  &5760.245    &   -37.9   &   1.2     &   2   \\
07/17/2011  &5760.398    &   -38.4   &   1.8     &   4   \\
07/20/2011  &5762.932    &   -34.8   &   1.4     &   6   \\
07/20/2011  &5763.372    &   -32.1   &   1.4     &   8   \\
07/20/2011  &5763.383    &   -35.4   &   2.0     &   4   \\
07/21/2011  &5764.440    &   -31.2   &   3.8     &   4   \\
07/22/2011  &5764.952    &   -32.2   &   0.7     &   2   \\
07/22/2011  &5765.450    &   -32.3   &   5.3     &   4   \\
07/23/2011  &5765.963    &   -31.7   &   1.2     &   2   \\
07/23/2011  &5766.362    &   -32.2   &   1.0     &   4   \\
07/23/2011  &5766.393    &   -29.7   &   0.8     &   8   \\
07/24/2011  &5766.974    &   -30.6   &   1.6     &   2   \\
07/25/2011  &5768.086    &   -29.8   &   1.1     &   2   \\
07/26/2011  &5769.085    &   -29.6   &   1.3     &   2   \\
07/29/2011  &5772.346    &   -27.3   &   0.7     &   5   \\
07/30/2011  &5773.360    &   -26.0   &   1.2     &   5   \\
08/01/2011  &5775.347    &   -24.2   &   0.7     &   5   \\
08/02/2011  &5776.345    &   -23.2   &   0.7     &   5   \\
08/03/2011  &5777.344    &   -22.2   &   1.4     &   5   \\
08/04/2011  &5778.356    &   -22.0   &   0.9     &   5   \\
08/05/2011  &5779.348    &   -20.8   &   1.2     &   5   \\
08/11/2011  &5784.968    &   -15.4   &   1.5     &   6   \\
08/12/2011  &5786.012    &   -16.1   &   1.8     &   6   \\
08/14/2011  &5788.024    &   -14.8   &   1.0     &   6   \\
\noalign{\smallskip}\hline \noalign{\smallskip}
\end{tabular}
\end{center}

\clearpage \noindent {\bf Table 1. continued}

\begin{center}
\begin{tabular}{rcrlc}
\hline\noalign{\smallskip}
Date        & JD2450000+    &  RV        & Error    & Source\\
MM/DD/YYYY  & days          &km\,s$^{-1}$&          &       \\
\noalign{\smallskip}\hline \noalign{\smallskip}
08/16/2011  &5790.228    &   -14.5   &   0.5     &   1   \\
08/19/2011  &5793.331    &   -10.5   &   3.4     &   5   \\
08/27/2011  &5800.985    &   -9.4    &   2.2     &   2   \\
08/28/2011  &5801.972    &   -9.4    &   2.0     &   2   \\
08/29/2011  &5803.065    &   -9.0    &   2.3     &   2   \\
08/30/2011  &5804.044    &   -10.7   &   2.5     &   2   \\
08/31/2011  &5805.001    &   -11.2   &   2.4     &   2   \\
09/01/2011  &5806.053    &   -12.9   &   3.5     &   2   \\
09/02/2011  &5807.025    &   -11.9   &   2.9     &   2   \\
09/04/2011  &5808.998    &   -11.0   &   1.9     &   2   \\
09/05/2011  &5809.963    &   -10.0   &   2.7     &   6   \\
09/05/2011  &5809.995    &   -12.0   &   1.8     &   2   \\
09/05/2011  &5810.023    &   -8.7    &   2.9     &   6   \\
09/06/2011  &5810.982    &   -11.4   &   2.6     &   2   \\
09/07/2011  &5811.990    &   -10.9   &   2.4     &   2   \\
10/02/2011  &5836.916    &   -10.1   &   1.2     &   6   \\
10/12/2011  &5846.913    &   -11.7   &   0.2     &   6   \\
\noalign{\smallskip}\hline
\end{tabular}
\end{center}

Column information: (1) -- Observing date, (2) -- Julian date, (3)
-- measured heliocentric radial velocity, (4) -- r.m.s. error of the
mirrored fit to the line profile, (5) -- source of the spectrum: 1
-- ESPaDOnS (CFHT), 2 -- FEROS at the 1.52m ESO telescope, 3 --
IAC80 telescope of the Teide Observatory, 4 -- C. Buil, 5 -- T.
Garrel, 6 -- B. Heathcote, 7 -- J. Ribeiro, 8 -- E. Pollmann, 9 --
O. Thizy.\\

\begin{table}
\caption[]{Radial velocities for the He {\sc ii} 4686 \AA\ line}
\label{t2}
\begin{center}
\begin{tabular}{rcrlc}
\hline\noalign{\smallskip}
Date        & JD2450000+    &  RV        & Error    & Source\\
MM/DD/YYYY  & days          &km\,s$^{-1}$&          &       \\
\noalign{\smallskip}\hline \noalign{\smallskip}
5/28/2010   &   5344.835    &   -10.7   &   0.0035  &   1  \\
5/30/2010   &   5346.835    &   -12.8   &   0.0014  &   1  \\
6/3/2010    &   5350.779    &   -7.7    &   0.0028  &   1  \\
6/4/2010    &   5351.802    &   -8.7    &   0.0017  &   1  \\
6/19/2010   &   5366.766    &   -10.7   &   0.0018  &   1  \\
6/22/2010   &   5369.747    &   -9.7    &   0.0019  &   1  \\
6/25/2010   &   5372.901    &   -13.8   &   0.0020  &   1  \\
7/22/2010   &   5399.741    &   -9.7    &   0.0019  &   1  \\
7/25/2010   &   5402.743    &   -9.7    &   0.0020  &   1  \\
7/27/2010   &   5404.763    &   -10.7   &   0.0026  &   1  \\
7/31/2010   &   5408.742    &   -9.7    &   0.0025  &   1  \\
8/5/2010    &   5413.842    &   -11.7   &   0.0030  &   1  \\
1/10/2011   &   5571.762    &   -20.5   &   0.0021  &   4  \\
1/16/2011   &   5577.729    &   -14.8   &   0.0039  &   4  \\
1/20/2011   &   5581.745    &   -14.2   &   0.0030  &   4  \\
1/23/2011   &   5584.749    &   -11.7   &   0.0034  &   4  \\
1/23/2011   &   5585.363    &   -11.8   &   0.0027  &   2  \\
2/9/2011    &   5601.725    &   -13.0   &   0.0027  &   4  \\
3/6/2011    &   5626.717    &   -19.4   &   0.0025  &   4  \\
3/7/2011    &   5628.349    &   -21.9   &   0.0033  &   2  \\
3/13/2011   &   5634.399    &   -14.0   &   0.0054  &   2  \\
3/16/2011   &   5637.392    &   -12.1   &   0.0038  &   2  \\
3/25/2011   &   5645.632    &   -18.0   &   0.0040  &   4  \\
4/16/2011   &   5667.567    &   -14.0   &   0.0029  &   4  \\
4/16/2011   &   5667.603    &   -8.2    &   0.0026  &   7  \\
4/16/2011   &   5668.391    &   -17.0   &   0.0038  &   2  \\
4/25/2011   &   5677.378    &   -6.9    &   0.0038  &   2  \\
4/28/2011   &   5679.559    &   -24.6   &   0.0030  &   4  \\
5/3/2011    &   5685.514    &   -13.8   &   0.0025  &   4  \\
5/11/2011   &   5692.535    &   -15.5   &   0.0031  &   4  \\
5/17/2011   &   5699.491    &   -17.1   &   0.0038  &   5  \\
\noalign{\smallskip}\hline \noalign{\smallskip}
\end{tabular}
\end{center}
\end{table}

\clearpage \noindent {\bf Table 2. continued}

\begin{center}
\begin{tabular}{rcrlc}
\hline\noalign{\smallskip}
Date        & JD2450000+    &  RV        & Error    & Source\\
MM/DD/YYYY  & days          &km\,s$^{-1}$&          &       \\
\noalign{\smallskip}\hline \noalign{\smallskip}
5/20/2011   &   5702.431    &   -23.2   &   0.0037  &   5   \\
5/21/2011   &   5703.053    &   -17.3   &   0.0019  &   6   \\
5/22/2011   &   5704.443    &   -21.5   &   0.0039  &   4   \\
5/24/2011   &   5706.427    &   -20.9   &   0.0031  &   5   \\
5/27/2011   &   5709.507    &   -18.8   &   0.0038  &   5   \\
5/28/2011   &   5709.573    &   -22.8   &   0.0025  &   4   \\
5/28/2011   &   5710.400    &   -24.2   &   0.0031  &   4   \\
6/7/2011    &   5721.026    &   -29.8   &   0.0028  &   1   \\
6/10/2011   &   5722.405    &   -25.5   &   0.0043  &   1   \\
6/12/2011   &   5724.565    &   -41.1   &   0.0039  &   7   \\
6/14/2011   &   5727.436    &   -28.1   &   0.0027  &   4   \\
6/14/2011   &   5727.446    &   -30.7   &   0.0039  &   5   \\
6/15/2011   &   5728.459    &   -29.7   &   0.0038  &   4   \\
6/17/2011   &   5729.268    &   -29.6   &   0.0023  &   1   \\
6/18/2011   &   5731.396    &   -43.9   &   0.0031  &   4   \\
6/18/2011   &   5731.406    &   -34.7   &   0.0031  &   5   \\
6/22/2011   &   5735.087    &   -39.7   &   0.0060  &   2   \\
6/25/2011   &   5738.984    &   -48.2   &   0.0044  &   2   \\
6/26/2011   &   5739.359    &   -42.9   &   0.0033  &   4   \\
6/29/2011   &   5742.396    &   -57.5   &   0.0035  &   3   \\
6/29/2011   &   5742.921    &   -55.8   &   0.0063  &   6   \\
7/1/2011    &   5743.551    &   -53.0   &   0.0027  &   3   \\
7/1/2011    &   5743.880    &   -49.9   &   0.0059  &   6   \\
7/1/2011    &   5744.374    &   -51.5   &   0.0023  &   4   \\
7/2/2011    &   5745.415    &   -57.2   &   0.0027  &   3   \\
7/2/2011    &   5745.480    &   -53.6   &   0.0029  &   4   \\
7/3/2011    &   5746.105    &   -56.6   &   0.0069  &   6   \\
7/4/2011    &   5746.938    &   -54.6   &   0.0047  &   6   \\
7/5/2011    &   5747.362    &   -52.6   &   0.0040  &   4   \\
7/6/2011    &   5748.419    &   -44.2   &   0.0260  &   3   \\
7/6/2011    &   5748.453    &   -51.5   &   0.0032  &   4   \\
7/6/2011    &   5749.122    &   -47.2   &   0.0040  &   2   \\
\noalign{\smallskip}\hline \noalign{\smallskip}
\end{tabular}
\end{center}

\clearpage \noindent {\bf Table 2. continued}
\begin{center}
\begin{tabular}{rcrlc}
\hline\noalign{\smallskip}
Date        & JD2450000+    &  RV        & Error    & Source\\
MM/DD/YYYY  & days          &km\,s$^{-1}$&          &       \\
\noalign{\smallskip}\hline \noalign{\smallskip}
7/9/2011    &   5751.354    &   -38.2   &   0.0027  &   4   \\
7/9/2011    &   5751.880    &   -37.9   &   0.0094  &   6   \\
7/10/2011   &   5752.358    &   -44.2   &   0.0029  &   4   \\
7/10/2011   &   5753.090    &   -46.9   &   0.0042  &   2   \\
7/11/2011   &   5753.360    &   -49.6   &   0.0031  &   4   \\
7/11/2011   &   5753.397    &   -35.9   &   0.0035  &   4   \\
7/11/2011   &   5754.085    &   -38.5   &   0.0037  &   2   \\
7/12/2011   &   5754.352    &   -48.0   &   0.0026  &   4   \\
7/12/2011   &   5755.976    &   -39.4   &   0.0043  &   2   \\
7/13/2011   &   5756.189    &   -38.6   &   0.0047  &   2   \\
7/14/2011   &   5756.880    &   -31.3   &   0.0029  &   5   \\
7/14/2011   &   5756.897    &   -29.0   &   0.0064  &   6   \\
7/14/2011   &   5757.211    &   -32.9   &   0.0058  &   2   \\
7/15/2011   &   5757.882    &   -29.4   &   0.0057  &   6   \\
7/17/2011   &   5759.961    &   -34.1   &   0.0048  &   2   \\
7/17/2011   &   5760.245    &   -34.3   &   0.0036  &   2   \\
7/18/2011   &   5760.373    &   -36.9   &   0.0021  &   4   \\
7/20/2011   &   5762.932    &   -26.1   &   0.0041  &   6   \\
7/21/2011   &   5763.363    &   -28.0   &   0.0026  &   4   \\
7/22/2011   &   5764.426    &   -30.3   &   0.0035  &   4   \\
7/22/2011   &   5764.952    &   -32.6   &   0.0035  &   2   \\
7/22/2011   &   5764.961    &   -31.2   &   0.0050  &   6   \\
7/23/2011   &   5765.963    &   -32.6   &   0.0037  &   2   \\
7/24/2011   &   5766.974    &   -28.6   &   0.0050  &   2   \\
7/25/2011   &   5768.086    &   -31.6   &   0.0039  &   2   \\
7/26/2011   &   5768.947    &   -27.4   &   0.0053  &   6   \\
7/26/2011   &   5769.085    &   -28.0   &   0.0038  &   2   \\
7/27/2011   &   5770.071    &   -25.8   &   0.0057  &   6   \\
7/28/2011   &   5770.969    &   -23.6   &   0.0052  &   6   \\
7/29/2011   &   5772.343    &   -25.9   &   0.0038  &   5   \\
\noalign{\smallskip}\hline \noalign{\smallskip}
\end{tabular}
\end{center}

\clearpage \noindent {\bf Table 2. continued}
\begin{center}
\begin{tabular}{rcrlc}
\hline\noalign{\smallskip}
Date        & JD2450000+    &  RV        & Error    & Source    \\
MM/DD/YYYY  & days          &km\,s$^{-1}$&          &           \\
\noalign{\smallskip}\hline \noalign{\smallskip}
7/31/2011   &   5774.007    &   -23.0   &   0.0061  &   6   \\
8/3/2011    &   5777.056    &   -28.2   &   0.0730  &   6   \\
8/3/2011    &   5777.334    &   -25.1   &   0.0041  &   5   \\
8/5/2011    &   5779.335    &   -23.3   &   0.0042  &   5   \\
8/11/2011   &   5784.953    &   -24.2   &   0.0056  &   6   \\
8/12/2011   &   5785.994    &   -15.5   &   0.0044  &   6   \\
8/14/2011   &   5788.023    &   -20.7   &   0.0080  &   6   \\
8/27/2011   &   5800.985    &   -24.4   &   0.0020  &   2   \\
8/28/2011   &   5801.972    &   -19.8   &   0.0037  &   2   \\
8/29/2011   &   5803.065    &   -16.1   &   0.0043  &   2   \\
8/30/2011   &   5804.044    &   -17.6   &   0.0038  &   2   \\
8/31/2011   &   5805.001    &   -18.3   &   0.0022  &   2   \\
9/1/2011    &   5806.053    &   -13.6   &   0.0034  &   2   \\
9/2/2011    &   5807.025    &   -15.1   &   0.0043  &   2   \\
9/4/2011    &   5808.998    &   -17.4   &   0.0025  &   2   \\
9/5/2011    &   5809.963    &   -17.9   &   0.0054  &   6   \\
9/5/2011    &   5809.995    &   -13.4   &   0.0025  &   2   \\
9/6/2011    &   5810.982    &   -12.6   &   0.0048  &   2   \\
9/7/2011    &   5811.990    &   -18.1   &   0.0037  &   2   \\
9/8/2011    &   5812.983    &   -19.6   &   0.0051  &   6   \\
10/2/2011   &   5836.916    &   -4.6    &   0.0049  &   6   \\
10/12/2011  &   5846.913    &   -5.1    &   0.0071  &   6   \\
\noalign{\smallskip} \hline
\end{tabular}
\end{center}

Column information: (1) -- Observing date, (2) -- Julian date, (3)
-- measured heliocentric radial velocity, (4) -- r.m.s. error of the
Gaussian fit to the line profile, (5) -- source of the spectrum: 1
-- ESPaDOnS (CFHT), 2 -- FEROS at the 1.52m ESO telescope, 3 --
IAC80 telescope of the Teide Observatory, 4 -- C. Buil, 5 -- T.
Garrel, 6 -- B. Heathcote, and 7 -- J. Ribeiro.\\

\begin{table}
\caption[]{Orbital solutions for $\delta$ Sco} \label{t3}
\begin{center}
\begin{tabular}{ccccccc}
\hline \noalign{\smallskip}
T$_0$     &  K$_{1}$    & $e$            & $\gamma$    & $\omega$   & Reduced    & N\\
days      &km\,s$^{-1}$ &                & km\,s$^{-1}$& degrees    & $\chi^{2}$ &  \\
\noalign{\smallskip} \hline \noalign{\smallskip}
2451797.9$\pm$1.0& 23.9$\pm$1.1& 0.937$\pm$0.002& $-5.8\pm$1.2& $-1.3\pm$3.5& 0.82&30\\
2455745.9$\pm$0.9& 23.4$\pm$0.5& 0.936$\pm$0.004& $-7.7\pm$0.8& $-2.6\pm$3.0& 0.74&25\\
2455745.9$\pm$1.0& 23.9$\pm$0.8& 0.936$\pm$0.003& $-6.6\pm$1.0& $-2.3\pm$3.8& 0.92&55\\
\noalign{\smallskip} \hline
\end{tabular}
\end{center}
\begin{list}{}
\item T$_0$ is the periastron passage epoch, K$_1$ is the
semi-amplitude of the radial velocity curve, $e$ is the orbit
eccentricity, $\gamma$ is the systemic radial velocity, $\omega$ is
the periastron longitude, N is the number of data points in a set.
The first line lists the best fit parameters for the periastron 2000
H$\alpha$ line radial velocity data set, the second line shows the
same for the 2011 He {\sc ii} 4686 \AA\ line, and the third line
shows the same for the combined data set.
\end{list}
\end{table}

\end{document}